\documentstyle[onecolumn,epsf]{mn}

\newif\ifAMStwofonts

\newcommand{\beq}{\begin{equation}}
\newcommand{\beqa}{\begin{eqnarray}}
\newcommand{\eeq}{\end{equation}}
\newcommand{\eeqa}{\end{eqnarray}}
\newcommand{\simg}{\ga}
\newcommand{\siml}{\la}
\newcommand{\bm}[1]{{\bmath{#1}}} 
\newcommand{\bme}{\hat {\bm e}}
\def\dddot#1{\stackrel{\cdots}{#1}}
\def\bari{\hbox{$I$\kern -.51em {\raise
      .53ex\hbox{$\scriptscriptstyle  - $}}}}
\title{Magnetic Deformation of Magnetars
for the Giant Flares of the Soft Gamma-Ray Repeaters}
\author[K. Ioka]
       {K. Ioka \\
        Department of Physics, Kyoto University, Kyoto 606-8502, Japan}
\date{Accepted .
      Received ;
      in original form 2000 March 21}

\pagerange{\pageref{firstpage}--\pageref{lastpage}}
\pubyear{2001}

\begin{document}

\maketitle

\label{firstpage}

\begin{abstract}
We present one possible mechanism for the giant flares
of the Soft Gamma-Ray Repeaters (SGRs) within the framework of magnetar,
i.e., superstrongly magnetized neutron star model,
motivated by the positive period increase associated with the August 27 event
from SGR 1900+14.
From the second-order perturbation analysis of the equilibrium of 
the magnetic polytrope, we find that there exist different equilibrium states
separated by the energy of the giant flares and the shift in the moment 
of inertia to cause the period increase.
This suggests that, if we assume that the global reconfiguration of 
the internal magnetic field of $H\simg 10^{16}$ G suddenly occurs,
the positive period increase $\Delta P_t/P_t \sim 10^{-4}$ as well as 
the energy $\simg 10^{44}$ ergs of the giant flares may be explained.
The moment of inertia can increase with a release of energy,
because the star shape deformed by the magnetic field can be prolate rather 
than oblate.
In this mechanism, since the oscillation of the neutron star will be excited,
a pulsation of $\sim$ ms period in the burst profile and 
an emission of the gravitational waves are expected.
The gravitational waves could be detected by the planned interferometers
such as LIGO, VIRGO and LCGT.
\end{abstract}


\section{INTRODUCTION}\label{sec:intro}
\subsection{Giant flares of SGRs}
Soft Gamma-ray Repeaters (SGRs) are persistent X-ray sources 
with X-ray luminosity $\sim 10^{35}$-$10^{36}$ erg s$^{-1}$ 
that emit brief ($\sim 0.1$ s),
intense ($\sim 10^{39}$-$10^{42}$ erg s$^{-1}$) recurrent bursts of
soft ($\sim 30$ keV) gamma-ray (see Hurley 1999 for a review).
There are four known SGRs, 
SGR 1900+14 (Mazets, Golenetskii \& Guryan 1979), 
SGR 1806-20 (Laros et al. 1986), 
SGR 1627-41 (Woods et al. 1999a) in the
Galactic plane and SGR 0525-66 (Golenetskii et al. 1979; Mazets et al. 1979;
Cline et al. 1980) 
in the Large Magellanic Cloud,
and one possible candidate, SGR 1801-23 (Cline et al. 2000)
towards the Galactic center.
Their apparent associations with supernova remnants
support that SGRs are neutron stars (Hurley 1999).

The spin periods and the spin-down rates of SGR 1900+14
(Woods et al. 1999b) and SGR 1806-20 (Kouveliotou et al. 1998)
are measured from the quiescent X-ray pulses.
If the spin-down is driven by magnetic dipole radiation,
the implied magnetic fields are $H \simg 10^{14}$ G,
which are greater than the quantum critical field
$H_{cr}=m_e^2 c^3/e \hbar \sim 4.4 \times 10^{13}$ G
(Baring \& Harding 1998; but see also Camilo et al. 2000).
Although the spin-down may not be driven by the dipole radiation
but a relativistic particle wind
(Thompson \& Blaes 1998; Kouveliotou et al. 1999;
Harding, Contopoulos \& Kazanas 1999; 
Rothschild, Marsden \& Lingenfelter 1999),
the rotational energy loss rate
($|{\cal I} \Omega \dot \Omega| \sim 10^{34}$ erg s$^{-1}$) is not sufficient 
for the persistent X-ray emission, the particle winds
and the soft gamma-ray bursts ($\simg 10^{36}$ erg s$^{-1}$),
and hence the superstrong magnetic field $H \simg 10^{14}$ G
is indispensable from energetics
if the energy source is the magnetic field.
Such superstrongly magnetized neutron stars are called ``magnetars''
(Duncan \& Thompson 1992; Thompson \& Duncan 1993, 1995, 1996),
and SGRs are better understood within the framework of magnetar 
than other possibilities, such as accretion models
(e.g., Kaplan et al. 2001; Liang 1995; Marsden et al. 1999; 
Cheng \& Dai 1998, 1999).
Anomalous X-ray pulsars (AXPs) are another group of X-ray sources
that are similar to the SGRs but have no burst activity
(see Mereghetti 1999 for a review).

During the past $\sim 20$ years,
two giant flares have been recorded
from two of the SGRs:
1979 March 5 event from SGR 0526-66 (Golenetskii et al. 1979;
Mazets et al. 1979; Cline et al. 1980)
and 1998 August 27 event from SGR 1900+14 (Hurley et al. 1999;
Feroci et al. 1999; Mazets et al. 1999).
These flares differ from the more common bursts in 
larger energies ($\simg 10^{44}$ ergs),
longer durations ($\simg 100$ s) and harder initial spectra ($\sim$ MeV).
In the August 27 event, the giant flare was intense
enough to ionize the Earth's night-side lower ionosphere
to the levels usually found only during daytimes (Inan et al. 1999).
A radio afterglow was also found (Frail, Kulkarni \& Bloom 1999).
In the magnetar model, it is considered that
the giant flares are produced by the global 
reconfiguration of the internal magnetic field,
while the smaller bursts are produced by ``crustquakes'' in the neutron star
(Thompson \& Duncan 1995).

Remarkably, during an 80 day interval in the summer of 1998
which contains the August 27 event,
the average spin-down rate of SGR 1900+14 increased by a factor $\sim 2.3$
(Woods et al. 1999b).
The sampling of the period history of SGR 1900+14
is insufficient to distinguish between a long-term
($\sim 80$ days) increase of the spin-down rate 
and a sudden increase in the spin period (a ``braking glitch'')
at the time of the August 27 event.\footnote{
Rothschild, Marsden \& Lingenfelter (1999) argue that
a sudden period increase at the August 27 event appears to be at odds
based on the relatively large period derivative
$\dot P_t \sim 1\times 10^{-10} {\rm s/s}$
measured about three months before the giant flare (Kouveliotou et al. 1999)
and just after the event (Murakami et al. 1999).
However a reanalysis of all {\it RXTE} observation
gives lower period derivatives,
although the same data are used (Woods et al. 1999).
Moreover, the timing noise of SGRs may be large (Woods et al. 2000)
so that the long term trend may be more important than the short term trend.
Therefore, we think that a sudden period increase is consistent with
the observation.}
In either case, this variable spin-down of SGR 1900+14 should
provide an important clue to the nature of the SGRs.

Thompson et al. (2000) considered the physical mechanisms
for the positive period increase of the observed magnitude
$\Delta P_t/P_t\sim 10^{-4}$ directly associated with the August 27 event.
They focused on two mechanisms: one is a loss of angular momentum 
by a particle wind and the other is an exchange of angular momentum
between the crustal neutron superfluid and the rest of the neutron star.
The former mechanism can induce the observed spin-down
only if an additional outflow of $\sim 10^{44}$ ergs 
lasts longer than the observed duration $\sim 10^{4}$ s
and shorter than $\sim 10^{5}$ s, and hence a fine tuning may be needed.
The latter mechanism may cause the observed spin-down.
However, at present, the exact cause for the positive period increase
of SGR 1900+14 and the giant flares remains unknown.

In this paper, we will present other mechanism for the giant flare
within the framework of magnetar,
in which the global reconfiguration of the internal magnetic field
of $H \simg 10^{16}$ G causes the giant flare of $\simg 10^{44}$ ergs
as well as the positive period increase of $\Delta P_t/P_t\sim 10^{-4}$.

\subsection{Order-of-magnitude estimate}\label{sec:order}
Before discussing the details,
we shall make an order-of-magnitude estimate.
Let us consider a rotating star.
There is a relation,
\beq
J={\cal I}\Omega,
\eeq
between the angular momentum $J$, the angular velocity $\Omega=2\pi/P_t$ 
and the moment of inertia ${\cal I}$ of the star.
There are two ways to cause the spin-down:
one way is to decrease the angular momentum
and the other is to increase the moment of inertia.
The mechanism considered by Thompson et al. (2000) is the former one.
Our mechanism for the spin-down is the latter one.
If the angular momentum $J$ is conserved,
the increase of the moment of inertia required 
for the positive period increase of SGR 1900+14
is of order $\Delta {\cal I}/{\cal I} \sim \Delta P_t/P_t \sim 10^{-4}$.
This corresponds to a rotational energy loss for the star
$|{\cal I}\Omega \Delta \Omega| \sim 10^{41}$ ergs,
which is much smaller than the energy of the giant flare.

Let us consider the internal magnetic field of $H \simg 10^{16}$ G.
This is somewhat larger than the external dipole fields
$H \sim 10^{14}$ G deduced from the period and the period derivative.
Such larger internal field $H \simg 10^{16}$ G is plausible 
if the magnetic field is generated 
by the helical dynamo in a new born neutron star
(Duncan \& Thompson 1992; Thompson \& Duncan 1993).
The magnetic field is a source of non-hydrostatic stress
in the interior of the star.
Hence the internal magnetic field induces
a deformation of the star, and
the magnetic deformation dominates the rotational one
when $H \simg 10^{14} (P_t/1 {\rm s})^{-1}$ G.
The induced difference in the moment of inertia of the star 
is given by order of
\beq
{{\Delta {\cal I}}\over{{\cal I}}} \sim \delta 
\sim {{(H^2/8\pi)(4\pi R^3/3)}\over{G M^2/R}}
\sim 10^{-4} \left({{H}\over{2 \times 10^{16} {\rm G}}}\right)^2
\left({{R}\over{10^6 {\rm cm}}}\right)^4
\left({{M}\over{1.4 M_{\odot}}}\right)^{-2},
\label{eq:deltaI}
\eeq
where $\delta$ is the ratio of the magnetic energy 
${\cal M}\sim (H^2/8\pi)(4\pi R^3/3) \sim 7 \times 10^{49} 
(H/2 \times 10^{16} {\rm G})^2 {\rm ergs}$
to the gravitational energy $|{\cal W}|\sim G M^2/R \sim 5 \times 10^{53}
{\rm ergs}$, and
$M$ and $R$ are the mass and the radius of the star
respectively (see below; see also Bocquet et al. 1995;
Bonazzola \& Gourgoulhon 1996; Konno, Obata \& Kojima 1999).
The deformation of the star also causes
the gravitational potential energy ${\cal W}\sim -G M^2/R$
to change by order of
(Chandrasekhar 1969; Shapiro \& Teukolsky 1983)
\beq
\Delta {\cal W} \sim \delta^2 |{\cal W}|/5 \sim 10^{45}
\left({{H}\over{2 \times 10^{16} {\rm G}}}\right)^4
\left({{R}\over{10^6 {\rm cm}}}\right)^7
\left({{M}\over{1.4 M_{\odot}}}\right)^{-2}
\quad {\rm ergs}.
\label{eq:deltaW}
\eeq
Here we assume that the deformation is elliptical like a rotating star.
Because the magnetic pressure is strongly anisotropic,
this is a good approximation for globally coherent fields,
such as a dipole poloidal field.
In this case, $\delta$ has the same meaning as the oblateness 
parameter $\epsilon$ in equations (10.11.2) and (10.11.7) 
of Shapiro \& Teukolsky (1983).
The energy shift is of order $\delta^2$
because the gravitational energy as a function of
the oblateness parameter $\delta$ is minimum at $\delta=0$.\footnote{
If the magnetic fields were well tangled,
the deformation would not be elliptical but radial
so that the energy shift is of order $\delta$.
In this case, our model gives too much energy.}
We will confirm that these crude estimates in equations 
(\ref{eq:deltaI}) and (\ref{eq:deltaW}) 
are good approximations in Section 4.2.

Therefore, if the global rearrangement of the internal magnetic field
$H \sim 10^{16}$ G occurs,
the moment of inertia and the energy
will change by order of $\Delta {\cal I}/{\cal I} \sim 10^{-4}$ and
$\Delta {\cal W} \sim 10^{45}$ ergs respectively,
which is comparable with the required values for the positive period increase
and the giant flare of SGR 1900+14.
In this way, the global reconfiguration of the internal magnetic field 
$H \simg 10^{16}$ G may explain the positive period increase 
$\Delta P_t/P_t \sim 10^{-4}$ as well as the energy 
$\simg 10^{44}$ ergs of the giant flare.
This coincidence of the order-of-magnitude estimate
and the observation may be by chance but is interesting enough to explore.
Note that the energy source of this mechanism is not the magnetic energy
but the gravitational energy.
Furthermore, the spin-down rate before and after the giant flare is 
almost the same since the moment of inertia and the external magnetic fields 
are nearly constant.

The first point that one may wonder is why the spin-down occurs
with a release of energy.
In a rotating star case, such as in the starquake model 
for the ordinary glitches (Ruderman 1969; Baym \& Pines 1971),
the energy release is always associated with the spin-up,
since a rotating star has less oblate equilibrium shape
as the star slows down.
Nevertheless, it is not a trivial problem whether or not 
the spin-down occurs with a release of energy
in a magnetized star case,
since it is possible for a magnetized star to be prolate rather than oblate.
Therefore, in this paper we shall address the following problem
as a first step: {\it can the spin-down occur with a release of energy
in a magnetized star ?}
To approach this problem, we consider the most idealized model.
We prepare several equilibrium polytropes with 
different magnetic configurations assuming axisymmetry and no rotations.
Then, we investigate the relation of the energy and the moment of inertia
between these equilibria
to find at least one example in which 
the moment of inertia increases with a release of energy.

In Section \ref{sec:equili}, we will study the equilibrium of a magnetized
polytropic star.
In Section \ref{sec:energy}, we will obtain the expression for the 
energy and the moment of inertia tensor.
In Section \ref{sec:examples}, we will compare equilibrium configurations
and show that the spin-down with a release of energy is possible.
We will also apply the results to the giant flares of SGRs.
Section \ref{sec:discuss} is devoted to discussions and summary.
We will consider observational signatures to test this model
in the future.

\section{THE EQUILIBRIUM OF MAGNETIC POLYTROPES}\label{sec:equili}
The equilibrium of a magnetized star has been studied
by several authors since the pioneering work of
Chandrasekhar and Fermi (1953).
If one assumes a barytropic equation of state,
one can make some progress in the understanding of the magnetic field
on the equilibrium configurations.
The equilibrium of a polytropic star with an
axisymmetric poloidal magnetic field
has been examined by Monaghan (1965, 1966a,b),
and has been treated in a systematic manner by Trehan \& Billings (1971).
The structure of the magnetic field with a toroidal and a poloidal
component in a polytropic star has been examined by Roxburgh (1966),
and the equilibrium of a polytropic star with a
toroidal and a poloidal magnetic field has been studied
in a systematic manner by Trehan \& Uberoi (1972).
These models are constructed using a perturbation technique
with the ratio of the magnetic energy to the gravitational energy
as the perturbation parameter.
This is essentially similar to that
developed by Chandrasekhar (1933) and Chandrasekhar \& Lebovitz (1962)
for a slowly rotating polytrope,
in which the perturbation parameter is the ratio of the rotational energy
to the gravitational energy.
Since the ratio $\delta$ of the magnetic energy
to the gravitational energy is small
as long as $H \siml 10^{18} (R/10^6 {\rm cm})^{-4} (M/M_{\odot})^{2}$ G, 
this method is the most suitable for our purposes.
As we can see from equation (\ref{eq:deltaW}), 
it is not sufficient to consider only 
the first-order in a small parameter $\delta$.
Therefore we will extend and generalize the previous works to the
second-order in a small parameter $\delta$.

\subsection{Basic equations}
The basic equations governing the hydrostatic equilibrium of a perfect
conducting fluid are
\beqa
0 &=& -\nabla p + \rho \nabla \Phi + 
{{1}\over{4 \pi}} (\nabla \bm \times \bm H) \bm \times \bm H,
\label{eq:eular}
\\
\nabla^2 \Phi &=& - 4 \pi G \rho,
\label{eq:laplace}
\\
\nabla \cdot \bm H &=& 0,
\label{eq:rotH}
\eeqa
where $p$ is the pressure, $\rho$ is the density, 
$\Phi$ is the gravitational potential
and $\bm H$ is the magnetic field.
We assume the polytropic equations of state,
\beq
p=K \rho^{1+1/n},
\label{eq:eos}
\eeq
and we normalize the quantities as
\beqa
\rho&=&\rho_c \Theta^n,\quad
r = \alpha \xi 
= \left[{{K(n+1) \rho_c^{-1+1/n}}\over{4\pi G}}\right]^{1/2} \xi,\quad
\bm H= (4 \pi G \delta)^{1/2} \rho_c \alpha \bm h,\quad
\Phi=4\pi G \alpha^2 \rho_c \phi,
\label{eq:norm}
\eeqa
where $\Theta$, $\xi$, $\bm h$ and $\phi$ are dimensionless quantities,
$n$ is the polytropic index, $\rho_c$ is the density at the center,
and $K$ is a constant.
Then, equations (\ref{eq:eular}), (\ref{eq:laplace}) and (\ref{eq:rotH})
are written as
\beqa
0 &=& -\nabla \Theta + \nabla \phi + 
\delta {{(\nabla \bm \times \bm h) \bm \times \bm h}\over{4 \pi \Theta^n}},
\label{eq:eular0}
\\
\nabla^2 \phi &=& - \Theta^n,
\label{eq:laplace0}
\\
\nabla \cdot \bm h &=& 0,
\label{eq:rotH0}
\eeqa
where $\delta$
represents, apart from a numerical factor, 
the ratio of magnetic to gravitational energy,
and $\nabla$ is the divergence with respect to $\xi$.

We shall assume that the magnetic field is axisymmetric about the $z$-axis.
Then we can express the magnetic field as a superposition
of a poloidal and a toroidal field in terms of two scalar functions
$P(\xi,\mu)$ and $T(\xi,\mu)$ in the manner
(e.g. Chandrasekhar 1961),
\beq
{{\bm h}\over{(4 \pi)^{1/2}}}=\nabla \bm \times (\varpi P {\bme}_{\varphi})
+\varpi T {\bme}_{\varphi},
\label{eq:H}
\eeq
where $\bme_{\varphi}$ is a unit vector along
the $\varphi$-direction and
$(\varpi, \varphi, z)$ denotes the cylindrical coordinate,
which is related to the spherical coordinate $(\xi, \theta, \varphi)$ 
as $\varpi=\xi \sin \theta=\xi (1-\mu^2)^{1/2}$
and $z=\xi \cos \theta=\xi \mu$.
Note that the projection of the lines of force on the meridional planes
gives $\varpi^2 P={\rm const}$.
Since equation (\ref{eq:eular0}) requires
$\nabla \bm \times \left[{(\nabla \bm \times \bm h)
\bm \times \bm h}/{\Theta^n}\right]=0$,
it can be shown that (Woltjer 1959; Ferraro 1954; Parker 1979)
\beqa
0&=&\nabla \left[\Theta - \phi - \delta N_P(\varpi^2 P)\right],
\label{eq:integ}
\\
\Delta_5 P&=&-\Theta^n N_P'(\varpi^2 P)-T N_T'(\varpi^2 P),
\label{eq:P}
\\
\varpi^2 T&=&N_T(\varpi^2 P),
\label{eq:T}
\eeqa
where $N_T(x)$ and $N_P(x)$ are arbitrary functions
which characterize the configuration of the magnetic fields, and
\beq
\Delta_5={{\partial^2}\over{\partial \varpi^2}}
+{{3}\over{\varpi}}{{\partial}\over{\partial \varpi}}
+{{\partial^2}\over{\partial z^2}}
={{\partial^2}\over{\partial \xi^2}} 
+{{4}\over{\xi}}{{\partial}\over{\partial \xi}}
+{{1-\mu^2}\over{\xi^2}}{{\partial^2}\over{\partial \mu^2}}
-{{4 \mu}\over{\xi^2}}{{\partial}\over{\partial \mu}},
\eeq
is the five-dimensional axisymmetric Laplacian.
Taking the divergence of equation (\ref{eq:integ}),
we obtain
\beq
\nabla^2 \Theta = -\Theta^n + \delta \nabla^2 N_P(\varpi^2 P),
\label{eq:theta}
\eeq
with equation (\ref{eq:laplace0}).

\subsection{Perturbative approach}
We shall suppose that the magnetic field is so small that $\delta$ 
may be treated as a small perturbation parameter.
On this assumption, we will obtain the solutions for 
the density $\Theta$, the poloidal field $P$ and the toroidal field $T$
to the second-order in $\delta$.
We assume that the solutions for $\Theta$, $P$ and $T$ have the forms,
\beqa
\Theta&=&\Theta_0+\delta\Theta_1+\delta^2\Theta_2+O(\delta^3),
\label{eq:thetaexp}
\\
P&=&P_0+\delta P_1+O(\delta^2),
\label{eq:Pexp}
\\
T&=&T_0+\delta T_1+O(\delta^2).
\label{eq:Texp}
\eeqa
By substituting these equations into the basic equations (\ref{eq:P}),
(\ref{eq:T}) and (\ref{eq:theta}), we obtain in each order of $\delta$
\beqa
\nabla^2 \Theta_0 &=& -\Theta_0^n,
\label{eq:theta0}
\\
\nabla^2 \Theta_1 &=& -n \Theta_0^{n-1} \Theta_1 + \nabla^2 N_P(\varpi^2 P_0),
\label{eq:theta1}
\\
\nabla^2 \Theta_2 &=& -n \Theta_0^{n-1} \Theta_2
-{{n(n-1)}\over{2}} \Theta_0^{n-2} \Theta_1^2 + 
\nabla^2 \left[\varpi^2 P_1 N_P'(\varpi^2 P_0)\right],
\label{eq:theta2}
\\
\Delta_5 P_0 &=& - \Theta_0^n N_P'(\varpi^2 P_0) - T_0 N_T'(\varpi^2 P_0),
\label{eq:P0}
\\
\Delta_5 P_1 &=& - n \Theta_0^{n-1} \Theta_1 N_P'(\varpi^2 P_0)
-\Theta_0^n \varpi^2 P_1 N_P''(\varpi^2 P_0)
-T_1 N_T'(\varpi^2 P_0) - T_0 \varpi^2 P_1 N_T''(\varpi^2 P_0),
\label{eq:P1}
\\
\varpi^2 T_0 &=& N_T(\varpi^2 P_0),
\label{eq:T0}
\\
\varpi^2 T_1 &=& \varpi^2 P_1 N_T'(\varpi^2 P_0).
\label{eq:T1}
\eeqa
These equations are to be solved order by order.
It is obvious that depending upon the choice
of the arbitrary functions $N_P(x)$ and $N_T(x)$, 
there can be many solutions
for the functions $P$ and $T$ which are consistent with 
the demand for equilibrium.
Therefore it is convenient to make general discussions
before specifying the functions $N_P(x)$ and $N_T(x)$.
So, we will consider the formal solutions for 
the density $\Theta$ and the gravitational potential 
$\phi$ in the next section, and
the formal expressions for the energy and the moment of inertia tensor
of the magnetic polytrope in Section \ref{sec:energy},
without solving the magnetic fields $P$ and $T$.
We will specify the magnetic fields $P$ and $T$ in Section \ref{sec:examples}.

\subsection{The solution for $\Theta$ and $\phi$ to second-order in $\delta$}
We will obtain the formal solutions for the density $\Theta$ and 
the gravitational potential $\phi$ before considering the 
specific magnetic fields.
Firstly we perform the Legendre expansion as 
\beqa
\Theta_1&=&\sum_{m=0}^{\infty} \psi_m(\xi) P_m(\mu),
\label{eq:theta1P}\\
\Theta_2&=&\sum_{m=0}^{\infty} \gamma_m(\xi) P_m(\mu),
\label{eq:theta2P}\\
N_P(\varpi^2 P_0)&=&\sum_{m=0}^{\infty} \Psi_m(\xi) P_m(\mu),
\label{eq:NpP}\\
\varpi^2 P_1 N_P'(\varpi^2 P_0)&=&\sum_{m=0}^{\infty} \Gamma_m(\xi) P_m(\mu),
\label{eq:Np'P}
\eeqa
where $P_m(\mu)$ are Legendre polynomials of order $m$.
Then, from equations (\ref{eq:theta0}), (\ref{eq:theta1}) and 
(\ref{eq:theta2}), we find that the radial functions 
$\Theta_0$, $\psi_m$ and $\gamma_m$ satisfy the equations
\beqa
{\cal D}_0 \Theta_0(\xi)&=&-\Theta_0^n(\xi),
\label{eq:LEeq}\\
{\cal D}_m (\psi_m - \Psi_m) &=& -n \Theta_0^{n-1} \psi_m,
\label{eq:psim}\\
{\cal D}_m (\gamma_m - \Gamma_m)&=& -n \Theta_0^{n-1} \gamma_m
-{{n(n-1)}\over{2}} \Theta_0^{n-2} \sum_{l=0}^{\infty} \psi_l Q_{lm},
\label{eq:gammam}
\eeqa
where
\beq
{\cal D}_m={{1}\over{\xi^2}}{{d}\over{d \xi}}
\left(\xi^2 {{d}\over{d\xi}}\right)
-{{m(m+1)}\over{\xi^2}},
\label{eq:Dm}
\eeq
and $Q_{lm}$ is defined by
\beq
P_l(\mu) \sum_{m=0}^{\infty} \psi_m P_m(\mu)=
\sum_{m=0}^{\infty} Q_{lm} P_m(\mu).
\label{eq:defQ}
\eeq
Note that
$Q_{l0}=\psi_l/(2l+1)$ and
$\Theta_0$ in equation (\ref{eq:LEeq}) is, of course, the Lane-Emden 
function of index $n$.
The radial functions $\Theta_0$, $\psi_m$ and $\gamma_m$
in equations (\ref{eq:LEeq}), (\ref{eq:psim}) and (\ref{eq:gammam})
are to be solved by imposing the boundary conditions
\beq
\Theta_0(0)=1,\quad \Theta_0'(0)=0,\quad
\psi_m(0)=0,\quad \psi_m'(0)=0,\quad
\gamma_m(0)=0,\quad \gamma_m'(0)=0,
\label{eq:bcs}
\eeq
since we shall let $\rho_c$ denote the central density.

With equations (\ref{eq:theta1P}), (\ref{eq:theta2P}), 
(\ref{eq:NpP}) and (\ref{eq:Np'P}),
the solution for $\phi$ can be obtained by integrating 
equation (\ref{eq:integ}) as
\beqa
\phi=C_0+\Theta_0
+\delta \left[C_{1;0}+\sum_{m=0}^{\infty} \left(\psi_m-\Psi_m\right) P_m(\mu)
\right]
+\delta^2 \left[C_{2;0}+\sum_{m=0}^{\infty} \left(\gamma_m-\Gamma_m\right)
P_m(\mu) \right],
\label{eq:phi}
\eeqa
where we write the integral constant as
$\phi_0=C_0+\delta C_{1;0}+\delta^2 C_{2;0}$.
The gravitational potential outside the polytrope
satisfies Laplace's equation $\nabla^2 \tilde \phi=0$,
where the tilde in $\tilde \phi$
distinguishes the exterior potential $\tilde \phi$
from the interior one $\phi$.
The solution of Laplace's equation which should be associated with
the interior solution (\ref{eq:phi}) is, clearly
\beqa
\tilde \phi={{K_0}\over{\xi}}+\delta\sum_{m=0}^{\infty}
{{K_{1;m}}\over{\xi^{m+1}}} P_m(\mu)
+\delta^2 \sum_{m=0}^{\infty} {{K_{2;m}}\over{\xi^{m+1}}} P_m(\mu).
\label{eq:phiout}
\eeqa

It remains to determine the various constants which occur in the solutions
for $\phi$ and $\tilde \phi$ by imposing on them the boundary conditions,
\beq
\left.\phi\right|_{\Xi(\mu)}=\left.\tilde \phi\right|_{\Xi(\mu)},
\quad
\left.{{\partial \phi}\over{\partial \xi}}\right|_{\Xi(\mu)}
=\left.{{\partial \tilde \phi}\over{\partial \xi}}\right|_{\Xi(\mu)},
\label{eq:bcphi}
\eeq
where the boundary of the polytrope $\Xi(\mu)$ is given by
\beq
\Xi(\mu)=\xi_0+\delta \sum_{m=0}^{\infty} q_m P_m(\mu)+
\delta^2 \sum_{m=0}^{\infty} t_m P_m(\mu).
\eeq
Here $\xi_0$ is the first zero of the Lane-Emden function.
The requirement that the density $\Theta$ vanishes on the boundary surface 
$\Theta(\Xi(\mu),\mu)=0$ determines the constants $q_m$ and $t_m$ as
\beqa
q_m&=&-{{\psi_m(\xi_0)}\over{\Theta_0'(\xi_0)}},
\\
t_m&=&-{{1}\over{\Theta_0'(\xi_0)}}
\left\{\gamma_m(\xi_0)-{{1}\over{\Theta_0'(\xi_0)}}
\sum_{l=0}^{\infty}\left[{{\psi_l(\xi_0)}\over{\xi_0}}+
\psi_l'(\xi_0)\right] Q_{lm}(\xi_0)
\right\},
\eeqa
where $Q_{lm}$ is defined in equation (\ref{eq:defQ}).
Then, evaluating $\phi$ and $\tilde \phi$ and their derivatives on $\Xi(\mu)$,
the boundary conditions in equation (\ref{eq:bcphi}) give
\beqa
K_0&=&-\xi_0^2 \Theta_0'(\xi_0),\quad
C_0=-\xi_0 \Theta_0'(\xi_0),
\nonumber\\
K_{1;m}&=&-{{1}\over{m+1}}\xi_0^{m+2}
\left[\psi_m'(\xi_0)-\Psi_m'(\xi_0)\right],
\nonumber\\
C_{1;0}&=&-\psi_0(\xi_0)+\Psi_0(\xi_0)
-\xi_0\left[\psi_0'(\xi_0)-\Psi_0'(\xi_0)\right],
\nonumber\\
K_{2;m}&=&-{{1}\over{m+1}}\xi_0^{m+2}
\left[\gamma_m'(\xi_0)-\Gamma_m'(\xi_0)
+{\cal Q}_m \right],
\nonumber\\
C_{2;0}&=&-\gamma_0(\xi_0)+\Gamma_0(\xi_0)-\xi_0\left[\gamma_0'(\xi_0)
-\Gamma_0'(\xi_0)+{\cal Q}_0\right],
\label{eq:phiconst}
\eeqa
and the additional boundary conditions 
for $\psi_m$ and $\gamma_m$ with $m\ge 1$,
\beqa
(m+1)\left[\psi_m(\xi_0)-\Psi_m(\xi_0)\right]
+\xi_0\left[\psi_m'(\xi_0)-\Psi_m'(\xi_0)\right]&=&0
\quad {\rm for}\quad  m \ge 1,
\label{eq:bcpsim}\\
(m+1)\left[\gamma_m(\xi_0)-\Gamma_m(\xi_0)\right]
+\xi_0\left[\gamma_m'(\xi_0)-\Gamma_m'(\xi_0)+{\cal Q}_m\right]
&=&0
\quad {\rm for}\quad  m \ge 1,
\label{eq:bcgammam}
\eeqa
where
\beq
{\cal Q}_m={{n \Theta_0^{n-1}(\xi_0)}\over{2 \Theta_0'(\xi_0)}}
\sum_{l=0}^{\infty} \psi_l(\xi_0) Q_{lm}(\xi_0),
\label{eq:defQm}
\eeq
and $Q_{lm}$ is defined in equation (\ref{eq:defQ}).
For $n<1$, this constant ${\cal Q}_m$ diverges 
since $\Theta_0(\xi_0)=0$, which means that
the expansion with respect to $\delta$ breaks down.\footnote{
Chandrasekhar's perturbation technique
breaks down near the surface of the polytrope
since $|\Theta_0|<|\delta \Theta_1|$ at $\xi\sim\xi_0$.
For a strict argument, 
some matching procedures may be necessary
as in the study of a slowly rotating star 
(Monaghan \& Roxburgh 1965; Smith 1975; Singh \& Singh 1984).
However, from the study of a slowly rotating star,
it seems that the perturbation technique
gives a sufficiently accurate answer
for, at least, integrated quantities such as mass (Anand 1968; James 1964).
So we leave strict arguments in the future.
}
Hereafter we will consider the case of $n \ge 1$.

Now we get the formal solution of the problem, i.e.,
we can calculate the density $\Theta$ and 
the gravitational potential $\phi$ if the
magnetic fields $\Psi_m$ and $\Gamma_m$ in equations (\ref{eq:NpP}) 
and (\ref{eq:Np'P}) are given.

\section{ENERGY AND MOMENT OF INERTIA TENSOR}\label{sec:energy}
In this section, we will obtain the formal expression 
for the differences in the total energy and the moment of inertia tensor
between two equilibrium configurations.
The energy and the moment of inertia tensor
will be obtained in the expanded forms with respect to the small parameter
$\delta$.

\subsection{Mass}
The total mass of the fluid is given by
\beq
M=\int_V \rho d^3r = \left[{{K(n+1)}\over{4\pi G}}\right]^{3/2}
\rho_c^{-{{1}/{2}}+{{3}/{2n}}} \int_V \Theta^n d^3\xi.
\label{eq:tmass}
\eeq
In the following discussions,
we wish to compare configurations of the equal mass.
Since the magnetic field causes the density $\Theta$ to change,
the equality in the masses can be achieved by an adjustment
in the central density,
\beq
\rho_c=\rho_0+\delta \rho_1 + \delta^2 \rho_2.
\label{eq:rhoexp}
\eeq
With equations (\ref{eq:thetaexp}), (\ref{eq:theta1P}), 
(\ref{eq:theta2P}) and (\ref{eq:rhoexp}),
the total mass in equation (\ref{eq:tmass}) can be expanded as
\beq
M=M_0+\delta M_1+\delta^2 M_2,
\label{eq:massexp}
\eeq
where
\beqa
M_0&=&\int_0^{\xi_0} \Theta_0^n \xi^2 d\xi,
\\
M_1&=&\int_0^{\xi_0} n \Theta_0^{n-1} \psi_0 \xi^2 d\xi 
+{{3-n}\over{2n}} {{\rho_1}\over{\rho_0}} M_0,
\\
M_2&=&
\int_0^{\xi_0} \left[n\Theta_0^{n-1} \gamma_0 + 
\sum_{m=0}^{\infty} {{n(n-1)}\over{2(2m+1)}}
\Theta_0^{n-2} (\psi_m)^2 \right]\xi^2 d\xi
+{{3-n}\over{2n}}{{\rho_1}\over{\rho_0}} M_1
+{{3-n}\over{2n}}\left[{{\rho_2}\over{\rho_0}}
-{{n+3}\over{4 n}}\left({{\rho_1}\over{\rho_0}}\right)^2\right] M_0
-\xi_0^2 {\cal Q}_0.
\eeqa
The mass is normalized by the unit
\beq
C_M=4\pi \rho_0 \alpha_0^3,
\label{eq:C_M}
\eeq
where
\beq
\alpha_0=\left[{{K(n+1) \rho_0^{-1+1/n}}\over{4\pi G}}\right]^{1/2},
\label{eq:alpha0}
\eeq
is the unit of the length.
Note that ${\cal Q}_0$ in $M_2$ comes from
the upper limit in the integral, which is not $\xi_0$ but $\Xi(\mu)$, i.e.,
the integral in equation (\ref{eq:tmass}) has to be performed
as $\int_V \Theta^n d^3\xi=2\pi\int_{-1}^{1}d\mu \int_0^{\Xi(\mu)} 
d\xi \xi^2 \Theta^n$.
The requirement of the equal mass is assured by $M_1=0$ and $M_2=0$,
which determine $\rho_1$ and $\rho_2$ as
\beqa
{{\rho_1}\over{\rho_0}}
&=&-{{2n}\over{3-n}}{{\psi_0'(\xi_0)-\Psi_0'(\xi_0)}\over{\Theta_0'(\xi_0)}},
\label{eq:rho1}
\\
{{\rho_2}\over{\rho_0}}
&=&-{{2n}\over{3-n}}{{\gamma_0'(\xi_0)-\Gamma_0'(\xi_0)+{\cal Q}_0}
\over{\Theta_0'(\xi_0)}}
+{{n(n+3)}\over{(3-n)^2}}\left[{{\psi_0'(\xi_0)-\Psi_0'(\xi_0)}
\over{\Theta_0'(\xi_0)}}\right]^2,
\label{eq:rho2}
\eeqa
where we use equations (\ref{eq:tn}), (\ref{eq:ptn-1}) and (\ref{eq:gtn-1}).

\subsection{Energy}
The total energy of this system is given by
(e.g., Chandrasekhar 1961; Woltjer 1959)
\beqa
{\cal E}={\cal M} + {\cal U} + {\cal W},
\label{eq:totalE}
\eeqa
where
\beqa
{\cal M}&=&{{1}\over{8\pi C_E}} \int_V |\bm H|^2 d^3r=
{{1}\over{4\pi}} \left({{\rho_c}\over{\rho_0}}\right)^{-1/2+5/2n}
\delta \int_V \left(-{{1}\over{2}} \varpi^2 P \Delta_5 P
+{{1}\over{2}} \varpi^2 T^2\right) d^3\xi,
\label{eq:magE}
\\
{\cal U}&=&{{n}\over{C_E}} \int_V p d^3r=
{{1}\over{4\pi}} \left({{\rho_c}\over{\rho_0}}\right)^{-1/2+5/2n}
\int_V \left({{n}\over{n+1}} \Theta^{n+1}\right) d^3\xi,
\label{eq:intE}
\\
{\cal W}&=& - {{1}\over{2 C_E}} \int_V \rho \Phi d^3r=
{{1}\over{4\pi}} \left({{\rho_c}\over{\rho_0}}\right)^{-1/2+5/2n}
\int_V \left(-{{1}\over{2}} \Theta^n \phi\right) d^3\xi,
\label{eq:gravE}
\eeqa
are the magnetic energy,
the internal energy,
and the gravitational energy respectively,
and we use equations (\ref{eq:eos}), (\ref{eq:norm})
and (\ref{eq:H}).
The energy is measured in the unit of
\beq
C_E=4\pi K(n+1) \rho_0^{1+1/n} \alpha_0^3,
\label{eq:C_E}
\eeq
where $\alpha_0$ is given by equation (\ref{eq:alpha0}).

\subsubsection{Magnetic energy}
With equations (\ref{eq:Pexp}), (\ref{eq:Texp}) and (\ref{eq:rhoexp}),
the total magnetic energy in equation (\ref{eq:magE}) can be expanded as
\beq
{\cal M}=\delta {\cal M}_1 + \delta^2 {\cal M}_2,
\eeq
where
\beqa
{\cal M}_1&=&{{1}\over{4\pi}}
\int_V \left(-{{1}\over{2}} \varpi^2 P_0 \Delta_5 P_0 
+{{1}\over{2}} \varpi^2 T_0^2 \right) d^3\xi,
\\
{\cal M}_2&=&{{1}\over{4\pi}}
\int_V \left(-{{1}\over{2}} \varpi^2 P_1 \Delta_5 P_0 
-{{1}\over{2}} \varpi^2 P_0 \Delta_5 P_1 +\varpi^2 T_1 T_0
\right) d^3\xi
+{{5-n}\over{2n}}{{\rho_1}\over{\rho_0}} {\cal M}_1.
\eeqa
By using equations 
(\ref{eq:rho1}), (\ref{eq:M1relation}) and (\ref{eq:M2relation}),
the above expressions for ${\cal M}_1$ and ${\cal M}_2$
can be reduced to
\beqa
{\cal M}_1&=&\int_0^{\xi_0} \Theta_0^n \Psi_0' \xi^3 d\xi,
\label{eq:M1}\\
{\cal M}_2&=&\int_0^{\xi_0} \Theta_0^n \Gamma_0' \xi^3 d\xi
-{{5-n}\over{3-n}}{{\psi_0'(\xi_0)-\Psi_0'(\xi_0)}\over{\Theta_0'(\xi_0)}}
\int_0^{\xi_0} \Theta_0^n \Psi_0' \xi^3 d\xi
+\int_0^{\xi_0} \sum_{m=0}^{\infty} {{n}\over{2m+1}} \Theta_0^{n-1} \psi_m
\Psi_m' \xi^3 d\xi.
\label{eq:M2}
\eeqa

\subsubsection{Internal energy}
With equations (\ref{eq:thetaexp}), (\ref{eq:theta1P}), 
(\ref{eq:theta2P}) and (\ref{eq:rhoexp}),
the internal energy in equation (\ref{eq:intE}) can be expanded as
\beq
{\cal U}={\cal U}_0 + \delta {\cal U}_1 + \delta^2 {\cal U}_2,
\eeq
where
\beqa
{\cal U}_0&=&
{{n}\over{n+1}} \int_0^{\xi_0} \Theta_0^{n+1} \xi^2 d\xi,
\\
{\cal U}_1&=&
\int_0^{\xi_0} n \Theta_0^n \psi_0 \xi^2 d\xi + 
{{5-n}\over{2n}} {{\rho_1}\over{\rho_0}} {\cal U}_0,
\\
{\cal U}_2&=&
\int_0^{\xi_0} \left[n \Theta_0^n \gamma_0 + \sum_{m=0}^{\infty}
{{n^2}\over{2(2m+1)}}
\Theta_0^{n-1} (\psi_m)^2\right] \xi^2 d\xi
+{{5-n}\over{2n}} {{\rho_1}\over{\rho_0}} {\cal U}_1
+{{5-n}\over{2n}} \left[{{\rho_2}\over{\rho_0}}
-{{n+5}\over{4n}}\left({{\rho_1}\over{\rho_0}}\right)^2
\right] {\cal U}_0.
\eeqa
By using equations 
(\ref{eq:rho1}), (\ref{eq:rho2}), (\ref{eq:tn+1}), (\ref{eq:ptn}) and 
(\ref{eq:gtn}),
the above expressions for ${\cal U}_0$, ${\cal U}_1$ and ${\cal U}_2$
can be reduced to
\beqa
{\cal U}_0&=&{{n}\over{5-n}} \xi_0^3 \left[\Theta_0'(\xi_0)\right]^2,
\label{eq:U0}\\
{\cal U}_1&=&-{{n}\over{3-n}}
\int_0^{\xi_0} \Theta_0^n \Psi_0' \xi^3 d\xi,
\label{eq:U1}\\
{\cal U}_2&=&-{{n}\over{3-n}}\Biggl\{
\int_0^{\xi_0} \Theta_0^n \Gamma_0' \xi^3 d\xi
-{{5-n}\over{3-n}}{{\psi_0'(\xi_0)-\Psi_0'(\xi_0)}\over{\Theta_0'(\xi_0)}}
\int_0^{\xi_0} \Theta_0^n \Psi_0' \xi^3 d\xi
\nonumber\\
&+&\int_0^{\xi_0} \sum_{m=0}^{\infty} {{n}\over{2(2m+1)}} \Theta_0^{n-1} 
\psi_m \left(\psi_m+ 2 \xi \psi_m'\right) \xi^2 d\xi
+{{1}\over{3-n}} \xi_0^3 \left[\psi_m'(\xi_0)-\Psi_m'(\xi_0)\right]^2
\Biggr\}.
\label{eq:U2}
\eeqa
Therefore, from equations (\ref{eq:M1}), (\ref{eq:M2}), 
(\ref{eq:U1}) and (\ref{eq:U2}),
there are following relations between the internal energy and the magnetic 
energy,
\beqa
{\cal U}_1&=&-{{n}\over{3-n}} {\cal M}_1,
\label{eq:UMrel1}\\
{\cal U}_2&=&-{{n}\over{3-n}} \left({\cal M}_2 + \widehat {\cal M}_2\right),
\label{eq:UMrel2}
\eeqa
where
\beqa
\widehat{\cal M}_2
&=&{{1}\over{3-n}} \xi_0^3 \left[\psi_0'(\xi_0)-\Psi_0'(\xi_0)\right]^2
+\int_0^{\xi_0} \sum_{m=0}^{\infty}
{{n}\over{2(2m+1)}} \Theta_0^{n-1} \psi_m
\left[\psi_m+2\xi \left(\psi_m'-\Psi_m'\right)\right]\xi^2 d\xi.
\label{eq:hatM2}
\eeqa
Note that $\widehat {\cal M}_2$ does not contain
the second-order quantities, such as $\gamma_m$ and 
$\Gamma_m$ in equations (\ref{eq:theta2P}) and (\ref{eq:Np'P}), 
but the non-linear terms
of the first-order quantities, such as $\psi_m$ and $\Psi_m$
in equations (\ref{eq:theta1P}) and (\ref{eq:NpP}).

\subsubsection{Gravitational energy}
With equations (\ref{eq:thetaexp}), (\ref{eq:theta1P}), 
(\ref{eq:theta2P}), (\ref{eq:phi}) and (\ref{eq:rhoexp}),
the gravitational energy in equation (\ref{eq:gravE}) can be expanded as
\beq
{\cal W}={\cal W}_0+\delta {\cal W}_1+\delta^2 {\cal W}_2,
\eeq
where
\beqa
{\cal W}_0&=&-{{1}\over{2}} \int_0^{\xi_0}
\Theta_0^n \left(C_0+\Theta_0\right) \xi^2 d\xi,
\\
{\cal W}_1&=&-{{1}\over{2}} \int_0^{\xi_0} 
\left[n \Theta_0^{n-1} \psi_0\left(C_0+\Theta_0\right)
+\Theta_0^n \left(C_{1;0}+\psi_0-\Psi_0\right)\right] \xi^2 d\xi
+{{5-n}\over{2n}} {{\rho_1}\over{\rho_0}} {\cal W}_0,
\\
{\cal W}_2&=&-{{1}\over{2}} \int_0^{\xi_0} \Biggl\{
\left[n \Theta_0^{n-1} \gamma_0 
+ \sum_{m=0}^{\infty} {{n(n-1)}\over{2(2m+1)}} \Theta_0^{n-2}
\left(\psi_m\right)^2 \right]
\left(C_0+\Theta_0\right) 
\nonumber\\
&+&n \Theta_0^{n-1} \left[\psi_0 C_{1;0} + 
\sum_{m=0}^{\infty} {{1}\over{2m+1}} \psi_m\left(\psi_m-\Psi_m\right)\right]
+\Theta_0^n \left(C_{2;0}+\gamma_0-\Gamma_0\right)
\Biggr\} \xi^2 d\xi
\nonumber\\
&+&{{5-n}\over{2n}} {{\rho_1}\over{\rho_0}} {\cal W}_1
+{{5-n}\over{2n}} \left[{{\rho_2}\over{\rho_0}}-{{n+5}\over{4n}}
\left({{\rho_1}\over{\rho_0}}\right)^2\right] {\cal W}_0
+{{1}\over{2}} C_0 \xi_0^2 {\cal Q}_0.
\eeqa
Note that ${\cal Q}_0$ in ${\cal W}_2$ comes from
the upper limit in the integral, which is not $\xi_0$ but $\Xi(\mu)$.
By using equations 
(\ref{eq:phiconst}), (\ref{eq:rho1}), (\ref{eq:rho2}), 
(\ref{eq:tn})-(\ref{eq:ptn2}), 
(\ref{eq:gtn-1})-(\ref{eq:gtn2}) and (\ref{eq:pP}),
the above expressions for ${\cal W}_0$, ${\cal W}_1$ and ${\cal W}_2$
can be reduced to
\beqa
{\cal W}_0=-{{3}\over{n}} {\cal U}_0,\quad
{\cal W}_1={{n}\over{3-n}} {\cal M}_1,\quad
{\cal W}_2={{n}\over{3-n}} {\cal M}_2+{{3}\over{3-n}} \widehat{\cal M}_2.
\label{eq:W012}
\eeqa

\subsubsection{Total energy}
For configurations in static equilibria,
the virial theorem for the hydromagnetics is given by
(Chandrasekhar 1961)
\beq
{\cal W}+{\cal M}+{{3}\over{n}}{\cal U}=0.
\label{eq:virial}
\eeq
We have shown this theorem explicitly
to the second-order in $\delta$
with equations (\ref{eq:UMrel1}), (\ref{eq:UMrel2}) and (\ref{eq:W012}).
By eliminating ${\cal W}+{\cal M}$ in equation (\ref{eq:totalE}),
the total energy can be expressed as
\beq
{\cal E}=-{{3-n}\over{n}}{\cal U}
=-{{3-n}\over{n}}{\cal U}_0 + 
\delta {\cal M}_1 + \delta^2 \left({\cal M}_2+\widehat {\cal M}_2\right),
\label{eq:Eexp}
\eeq
with equations (\ref{eq:UMrel1}) and (\ref{eq:UMrel2}).
This equation (\ref{eq:Eexp}) can be interpreted as that the total energy
is the sum of the zeroth-order 
background energy ${\cal E}_0=-[(3-n)/n]{\cal U}_0$
and the total magnetic energy 
${\cal M}=\delta {\cal M}_1+\delta^2 {\cal M}_2$
with a correction $\delta^2 \widehat {\cal M}_2$
which comes from the product of the first-order quantities.

\subsection{Moment of inertia tensor}
The moment of inertia tensor is
diagonal in the chosen representation.
In view of the axisymmetry, the two distinct 
components of this tensor are\footnote
{The moment of inertia tensor can be constructed from
the second moment of the mass distribution $I_{ij}$
by ${\cal I}_{ij}=\delta_{ij} {\rm tr}(I_{ab})-I_{ij}$.
The $i$-component of the angular momentum is given by
$J_i={\cal I}_{ij} \Omega_j$ where $\Omega_j$ is the $j$-component of the 
angular velocity of the rotation.
(Of course, our analysis neglects the rotation.)}
\beqa
{\cal I}_{11}&=&{\cal I}_{22}=
{{1}\over{2 C_I}}\int_V \rho r^2 (1+\mu^2) d^3r =
{{1}\over{8\pi}} \left({{\rho_c}\over{\rho_0}}\right)^{-3/2+5/2n}
\int_V \Theta^n \xi^2 (1+\mu^2) d^3\xi,
\label{eq:I11}\\
{\cal I}_{33}&=&{{1}\over{C_I}}\int_V \rho r^2 (1-\mu^2) d^3r =
{{1}\over{4\pi}} \left({{\rho_c}\over{\rho_0}}\right)^{-3/2+5/2n}
\int_V \Theta^n \xi^2 (1-\mu^2) d^3\xi,
\label{eq:I33}
\eeqa
where the moment of inertia tensor is measured in the unit of
\beq
C_I=4\pi \rho_0 \alpha_0^5.
\eeq
With equations (\ref{eq:thetaexp}), (\ref{eq:theta1P}), 
(\ref{eq:theta2P}) and (\ref{eq:rhoexp}),
we can expand the moment of inertia tensor as
\beqa
{\cal I}_{11}&=&{\cal I}_{0}+\delta {\cal I}_{11;1},
\label{eq:Ixexp}
\\
{\cal I}_{33}&=&{\cal I}_{0}+\delta {\cal I}_{33;1},
\label{eq:Izexp}
\eeqa
where
\beqa
{\cal I}_{0}&=&{{2}\over{3}}
\int_0^{\xi_0} \Theta_0^n \xi^4 d\xi,
\label{eq:I11;0}\\
{\cal I}_{11;1}&=&{{2}\over{3}}
\left[\int_0^{\xi_0} n \Theta_0^{n-1} \left(\psi_0+{{1}\over{10}}\psi_2\right)
\xi^4 d\xi\right]
+{{5-3n}\over{2n}}{{\rho_1}\over{\rho_0}} {\cal I}_{0},
\label{eq:I11;1}\\
{\cal I}_{33;1}&=&{{2}\over{3}}
\left[\int_0^{\xi_0} n \Theta_0^{n-1} \left(\psi_0-{{1}\over{5}}\psi_2\right)
\xi^4 d\xi\right]
+{{5-3n}\over{2n}}{{\rho_1}\over{\rho_0}} {\cal I}_{0}.
\label{eq:I33;1}
\eeqa
As we will see later, it is sufficient for our discussions
to expand the moment of inertia tensor to the first-order
in $\delta$.

\subsection{Differences in energy and moment of inertia tensor
between equilibria}
We shall obtain the formal expressions for the differences
in energy and moment of inertia tensor between two equilibria
which have different configurations of the magnetic fields.
In order to compare two equilibria,
it is not sufficient to fix the mass.
We need two relations of parameters between equilibria
because there are two free parameters,
the central density $\rho_0$ and 
the ratio of the magnetic energy to the gravitational energy $\delta$,
in the magnetized polytropic model.
We shall take the magnetic energy as one more fixed parameter.
Let us characterize the configuration of the magnetic field
by $\lambda$, and consider that
the magnetic configuration changes from $\lambda_i$ to $\lambda_f$.
Hereafter let the subscript $i$ ($f$) mean the initial (final) state.
If we assume that the magnetic energy decays by some amount $\Delta {\cal M}$ 
in this transition, the relation of the magnetic energy 
between equilibria is given by
\beq
{\cal M}(\lambda_f)-{\cal M}(\lambda_i)=
\left[\delta_f {\cal M}_1(\lambda_f)+\delta_f^2 {\cal M}_2(\lambda_f)\right]
-\left[\delta_i {\cal M}_1(\lambda_i)+\delta_i^2 {\cal M}_2(\lambda_i)\right]
\equiv-\Delta {\cal M},
\label{eq:delM}
\eeq
which determines the relation between $\delta_i$ and $\delta_f$.
Since the amount of the decaying magnetic energy depends on the
process of the magnetic reconfiguration,
we will treat $\Delta {\cal M}$ as a parameter in our discussions.
Then we find from equations (\ref{eq:Eexp}) and (\ref{eq:delM}) that 
the difference in the total energy between two configurations
$\lambda_i$ and $\lambda_f$ is given by
\beqa
{{\Delta{\cal E}}\over{|{\cal E}_0|}}\equiv
{{{\cal E}(\lambda_f)-{\cal E}(\lambda_i)}\over{|{\cal E}_0|}}
=\delta_i^2 {{{\cal M}_1^2(\lambda_i)}\over{|{\cal E}_0}|}
\left[
{{\widehat{\cal M}_2(\lambda_f)}\over{{\cal M}_1^2(\lambda_f)}}
-{{\widehat{\cal M}_2(\lambda_i)}\over{{\cal M}_1^2(\lambda_i)}}
-{{\Delta {\cal M}}\over{\delta_i^2 {\cal M}_1^2(\lambda_i)}}
\right],
\label{eq:delE}
\eeqa
where ${\cal E}_0$ is the lowest order of the total energy,
${\cal E}_0=-[(3-n)/n]{\cal U}_0$, and 
we keep the lowest order of terms which include $\Delta {\cal M}$.
The energy is released when $\Delta {\cal E}<0$.
If we can neglect the term of $\Delta {\cal M}$,
i.e., $|\Delta {\cal M}| \siml \delta^2$,
the released energy is of order 
$|\Delta {\cal E}/{\cal E}_0| \sim \delta^2$.

From equations (\ref{eq:Ixexp}), (\ref{eq:Izexp}) and (\ref{eq:delM}),
the difference in the moment of inertia tensor between two configurations
$\lambda_i$ and $\lambda_f$ is given by
\beq
{{\Delta {\cal I}_{kk}}\over{{\cal I}_{0}}}
\equiv
{{{\cal I}_{kk}(\lambda_f)-{\cal I}_{kk}(\lambda_i)}\over{{\cal I}_{0}}}
=\delta_i {{{\cal M}_1(\lambda_i)}\over{{\cal I}_{0}}}
\left[
{{{\cal I}_{kk;1}(\lambda_f)}\over{{\cal M}_1(\lambda_f)}}
-{{{\cal I}_{kk;1}(\lambda_i)}\over{{\cal M}_1(\lambda_i)}}
-{{\Delta {\cal M}}\over{\delta_i {\cal M}_1(\lambda_i)}}
{{{\cal I}_{kk;1}(\lambda_f)}\over{{\cal M}_1(\lambda_f)}}
\right],
\label{eq:delI}
\eeq
keeping the lowest order in $\delta$.
The moment of inertia tensor increases when $\Delta {\cal I}_{kk}>0$.
If we can neglect the term of $\Delta {\cal M}$,
i.e., $|\Delta {\cal M}| \siml \delta$,
the difference in the moment of inertia tensor between two equilibria
is of order $|\Delta {\cal I}_{kk}/{\cal I}_{0}| \sim \delta$.

From equations (\ref{eq:delE}) and (\ref{eq:delI}), 
the conditions for the moment of inertia tensor 
to increase with a release of energy,
$\Delta {\cal E}<0$ and $\Delta {\cal I}_{kk}>0$,
are reduced to
\beqa
{{\widehat {\cal M}_2(\lambda_i)}\over{{\cal M}_1^2(\lambda_i)}}
+{{\eta}\over{\delta_i {\cal M}_1(\lambda_i)}}
&>&{{\widehat {\cal M}_2(\lambda_f)}\over{{\cal M}_1^2(\lambda_f)}},
\label{eq:Econd}
\\
{{1}\over{1-\eta}}{{{\cal I}_{kk;1}(\lambda_i)}\over{{\cal M}_1(\lambda_i)}}
&<&{{{\cal I}_{kk;1}(\lambda_f)}\over{{\cal M}_1(\lambda_f)}},
\label{eq:Icond}
\eeqa
where
\beq
\eta={{\Delta {\cal M}}\over{\delta_i {\cal M}_1(\lambda_i)}}<1,
\label{eq:etadef}
\eeq
is the ratio of the decaying magnetic energy 
to the initial total magnetic energy.
Note that these conditions depend on the zeroth- and the first-order quantities
and not on the second-order quantities,
i.e., ${\cal M}_1$, $\widehat {\cal M}_2$ and ${\cal I}_{kk;1}$
do not depend on $\gamma_m$ and $\Gamma_m$ but on
$\Theta_0$, $\psi_m$ and $\Psi_m$ 
as we can see from equations (\ref{eq:M1}),
(\ref{eq:hatM2}), (\ref{eq:I11;1}) and (\ref{eq:I33;1}).

\section{COMPARISON OF EQUILIBRIUM CONFIGURATIONS}\label{sec:examples}
So far we have obtained the general solutions and expressions
for the equilibrium configurations without solving magnetic fields.
In this section, we specify magnetic configurations
to investigate the conditions
for the moment of inertia tensor 
to increase with a release of energy in equations (\ref{eq:Econd})
and (\ref{eq:Icond}).
We shall obtain the magnetic fields to the lowest order in $\delta$,
i.e., $P_0$ and $T_0$, since the conditions 
in equation (\ref{eq:Econd}) and (\ref{eq:Icond}) 
do not depend on the second-order quantities
as noted in the last of the previous section.

\subsection{The solution for $P$ and $T$ to first-order in $\delta$}
\label{sec:TPmag}
We shall consider a special case of the magnetic fields characterized by
\beqa
N_P(x)=-x,\quad N_T(x)=\lambda x,
\label{eq:NpNt}
\eeqa
in which case the solutions for $P_0$ and $T_0$ are functions of only 
the radial coordinate $\xi$
and are easily calculated (Woltjer 1960; Trehan \& Uberoi 1972).
Then $\lambda$ controls the configurations of the magnetic fields.
Substituting the above expressions for $N_P(x)$ and $N_T(x)$
to equations (\ref{eq:P0}) and (\ref{eq:T0}),
we get equations to be solved,
\beqa
\Delta_5 P_0 + \lambda^2 P_0=\Theta_0^n,
\label{eq:P0spec}
\quad
T_0=\lambda P_0,
\label{eq:T0spec}
\eeqa
where $\Theta_0$ is the Lane-Emden function of index $n$.
Since we are considering the larger interior fields
compared with the exterior fields,
the most convenient way to impose the boundary condition
is to require that the fields vanish at the boundary
(Trehan \& Uberoi 1972), that is
to the lowest order in $\delta$,
\beq
P_0(\xi_0)=0,\quad {{\partial P_0}\over{\partial \xi}}(\xi_0)=0.
\label{eq:bcP0}
\eeq
Then, the solution for $P_0$ is given by 
(Woltjer 1960; Trehan \& Uberoi 1972)
\beqa
P_0(\xi)={{\lambda}\over{\xi}} n_1(\lambda \xi)
\int_0^{\xi} \Theta_0^n(x) j_1(\lambda x) x^3 dx
+{{\lambda}\over{\xi}} j_1(\lambda \xi)
\int_{\xi}^{\xi_0} \Theta_0^n(x) n_1(\lambda x) x^3 dx,
\eeqa
where $\lambda$ is a root of
\beq
\int_0^{\xi_0} \Theta_0^n(x) j_1(\lambda x) x^3 dx=0,
\label{eq:bclambda}
\eeq
and $j_m(\xi)$ and $n_m(\xi)$ are spherical Bessel and Neumann functions 
of order $n$ respectively.
The first few roots of equation (\ref{eq:bclambda}) are given 
in Table 1 for the polytropic indices $n=1, 1.5, 2$ and $2.5$.

Having determined the solution for $P_0$, we can calculate
$\Psi_m$ in equation (\ref{eq:NpP}) by
\beq
\Psi_0=-{{2}\over{3}} \xi^2 P_0,
\quad
\Psi_2={{2}\over{3}} \xi^2 P_0,
\quad
\Psi_m=0 \quad {\rm for}\quad  m\ne 0,2.
\eeq
Then the radial functions $\psi_m$ can be calculated
by solving equation (\ref{eq:psim})
subject to the boundary conditions in equations (\ref{eq:bcs}) and 
(\ref{eq:bcpsim}).
The quantities, ${\cal M}_1$, $\widehat {\cal M}_2$
and ${\cal I}_{kk;1}$, can be calculated\footnote{
In the numerical calculation, we can solve all differential equations 
simultaneously using the fifth-order Runge-Kutta method (Press et al. 1992).
}
from 
$\Theta_0$, $\psi_m$ and $\Psi_m$
by using equations (\ref{eq:M1}), (\ref{eq:hatM2}), (\ref{eq:I11;1}) 
and (\ref{eq:I33;1}).

Now, we are in a position to examine the conditions
for the moment of inertia tensor to increase with a release of energy
in equations (\ref{eq:Econd}) and (\ref{eq:Icond}).

\subsection{Results}\label{sec:result}
For the moment, we neglect the decaying magnetic energy, i.e., $\eta=0$
in equation (\ref{eq:etadef}).
Note that $\eta$ is a free parameter in our analysis
since we need two relations of parameters
(here the mass and the magnetic energy) between equilibria.
We evaluate $\widehat {\cal M}_2/{\cal M}_1^2$ (energy)
and ${\cal I}_{kk;1}/{\cal M}_1$ (moment of inertia)
by taking various values of $\lambda$
to examine the conditions for the moment of inertia tensor to increase 
with a release of energy in equations (\ref{eq:Econd}) and (\ref{eq:Icond}).
In Fig.~1, we plot $\widehat {\cal M}_2/{\cal M}_1^2$
and ${\cal I}_{kk;1}/{\cal M}_1$ as a function of $\lambda$
for the cases of $n=1$ and $n=2.5$.
$\lambda$ takes discrete values in Table 1.
We can see that the conditions in equations (\ref{eq:Econd}) and 
(\ref{eq:Icond}) are satisfied for the moment of inertia tensor
of $33$-component, ${\cal I}_{33;1}$, i.e.,
if the magnetic configuration
changes from the state of larger $\lambda$ to that of smaller $\lambda$,
$\widehat {\cal M}_2/{\cal M}_1^2$ (energy) decreases
while ${\cal I}_{33;1}/{\cal M}_1$ (moment of inertia) increases.
This means that the total energy decreases while
the moment of inertia tensor along the magnetic axis increases
from equations (\ref{eq:delE}) and (\ref{eq:delI}).
Therefore the answer for the question posed in Section \ref{sec:order}
is that the spin-down with a release of energy is possible 
in a magnetized star.

Note that the configuration of the polytrope is prolate
rather than oblate since the moment of inertia tensor
of $11$-component ${\cal I}_{11;1}$ is larger than
that of $33$-component ${\cal I}_{33;1}$ from Fig.~1.
Therefore, as the total energy decreases,
the configuration of the polytrope becomes more spherical.
This may meet our intuitions from the view
that the self-gravitating fluid is the most stable 
at the spherical configuration.

In Fig.~2, the projection of the lines of force on the meridional
plane, which is given by $\delta^{1/2} \varpi^2 P_0={\rm const}$,
is shown for the case of $n=1$ 
(see equations (4.17) and (6.17) of Parker 1979).
The lines of force lie in the two dimensional surfaces which are made by
rotating each loop in Fig.~2 about the symmetrical axis.
For $\lambda=\lambda_2$ the inner projected lines rotate 
in the different direction to the outer ones on the meridional plane.
The lines of force for larger $\lambda$ have the more
complicated structure than that for smaller $\lambda$,
i.e., the number of the smaller structure like the inner loops for 
$\lambda=\lambda_2$ increases as $\lambda$ increases.
Note that the flux of lines of force through 
any slice of the star is always zero
because of the boundary condition ${\bm h}(\Xi(\mu),\mu)=0$.

To evaluate physical quantities, we adopt
the canonical values,
$C_M M_0=1.4 M_{\odot}$ and $R=\alpha_0 \xi_0=10^{6}$ cm,
for the total mass and the radius of the polytrope respectively,
which determine $\rho_0$ and $K$ from equations (\ref{eq:C_M}) and 
(\ref{eq:alpha0}).
Furthermore we adopt $\Delta {\cal I}_{33}/{\cal I}_{0}=10^{-4}$
since this value is required for the positive period increase of SGR 1900+14.
Then we can calculate $\delta_i$ from equation (\ref{eq:delI}),
and hence the released energy $C_E \Delta {\cal E}$ 
from equation (\ref{eq:delE}),
where $C_E$ is the unit of the energy in equation (\ref{eq:C_E}).
In Fig.~3, we plot the released energy $C_E \Delta {\cal E}$ 
as a function of 
the various sets of the initial and final state,
$(\lambda_i,\lambda_f)\equiv10(f-1)-f(f-1)/2+i-f$,
where $f=1,\cdots,10,\quad i=f+1,\cdots,10$ and $f<i$
(see Table 2).
We can also obtain the magnetic energy ${\cal M}\simeq \delta {\cal M}_1$
required for $\Delta {\cal I}_{33}/{\cal I}_{0}=10^{-4}$
from equation (\ref{eq:delI}).
In Fig.~4, we plot the typical magnetic field $\bar H$,
which is defined by $({\bar H}^2/8 \pi)(4\pi R^3/3)=C_E {\cal M}$,
as a function of 
the various sets of the initial and final state,
$(\lambda_i,\lambda_f)\equiv10(f-1)-f(f-1)/2+i-f$ (see Table 2).
These results in Fig.~3 and 4 are
essentially the same as the 
order-of-magnitude argument in Section \ref{sec:order}.
Note that there are relations, $C_E \Delta {\cal E} \propto
(\Delta {\cal I}_{33}/{\cal I}_{0})^2$
and ${\bar H} \propto (\Delta {\cal I}_{33}/{\cal I}_{0})^{1/2}$.

When the decaying magnetic energy is not zero, $\eta \ne 0$,
it can occur that
the conditions for the moment of inertia tensor to increase 
with a release of energy in equations (\ref{eq:Econd}) and (\ref{eq:Icond})
are not satisfied.
In Fig.~5, we obtain the region of $\eta$ where
the conditions in equations (\ref{eq:Econd}) and (\ref{eq:Icond})
are satisfied as a function of
the various sets of the initial and final state,
$(\lambda_i,\lambda_f)\equiv10(f-1)-f(f-1)/2+i-f$,
where $f=1,\cdots,10,\quad i=f+1,\cdots,10$ and $f<i$ (see Table 2).
Here we assume $\Delta {\cal I}_{33}/{\cal I}_{0}=10^{-4}$ 
and consider the $33$-component of the 
moment of inertia tensor.
From this Figure, we find that
the conditions in equations (\ref{eq:Econd}) and (\ref{eq:Icond})
are satisfied when $-10^{-4} \siml \eta \siml 1$ for $n=1$
and $-10^{-4} \siml \eta \siml 10^{-3}$ for $n=2.5$.
Note that
the decaying magnetic energy is comparable with the released energy
when $|\eta| \sim 10^{-4} \sim \Delta {\cal I}_{33}/{\cal I}_{0}$
since $|\Delta {\cal M}| \sim |\eta {\cal M}|
\sim \delta |({\cal M}/{\cal E}) {\cal E}| \sim \delta^2 |{\cal E}|$.
Therefore the moment of inertia increases with a release of energy
if the decaying magnetic energy is less than the released energy.

The numerical values of the various parameters
are given in Tables 3 and 4.
We can calculate such as $\Delta {\cal E}$ and 
$\Delta {\cal I}_{kk}/{\cal I}_{0}$
using these Tables and equations (\ref{eq:delE}) and (\ref{eq:delI}).
We also give the ratio of the energies of the toroidal and 
the poloidal components of the magnetic field, 
${\cal M}_{T1}/{\cal M}_{P1}$.
To the lowest order in $\delta$, it is given by
\beqa
{\cal M}_{T1}&=&{{1}\over{4\pi}}\int_V \left({{1}\over{2}}
\varpi^2 T_0^2\right) d^3\xi=
{{\lambda^2}\over{3}}\int_0^{\xi_0}
P_0^2 \xi^4 d\xi,
\\
{\cal M}_{P1}&=&{{1}\over{4\pi}}\int_V \left(-{{1}\over{2}}
\varpi^2 P_0 \Delta_5 P_0\right) d^3\xi=
{\cal M}_{T1}-{{1}\over{3}}\int_0^{\xi_0} 
\Theta_0^n P_0 \xi^4 d\xi.
\eeqa
From Table 4, the polytrope becomes more spherical 
as the ratio ${\cal M}_{T1}/{\cal M}_{P1}$ decreases,
which is a consequence of the fact that toroidal fields tend to make 
the star prolate and poloidal fields tend to make it oblate.
It is interesting to note that the internal magnetic field
generated by a post-collapse $\alpha$-$\Omega$ dynamo
is probably dominated by a toroidal component (Duncan \& Thompson 1992;
Thompson \& Duncan 1993).

\section{DISCUSSIONS}\label{sec:discuss}
In this paper, we have proposed one possible mechanism for the giant flares
of the SGRs within the framework of magnetar,
motivated by the positive period increase associated with the August 27 event.
We assume that the global reconfiguration of the internal magnetic field
suddenly occurs.
Then, the shape of the magnetar is deformed and
this deformation causes the moment of inertia and the total energy
to change.
With the internal magnetic field of $H\simg 10^{16}$ G,
we can explain the positive period increase $\Delta P_t/P_t \sim 10^{-4}$
as well as the energy $\simg 10^{44}$ ergs of the giant flare of SGR 1900+14
if the magnetic field does not decay so much in this giant flare.
In this mechanism, the energy source is not the magnetic energy
but the gravitational energy.
In order to investigate whether or not 
the spin-down occurs with a release of the energy in a magnetized star,
we have analyzed the magnetic polytrope as the most idealized model
using the second-order perturbation technique.
We have found that it is possible for
the moment of inertia along the axisymmetric axis
to increase with a release of energy.
This result can be understood as follows.
The polytrope is prolate at the first state.
As the polytrope becomes more spherical, the total energy decreases
since the spherical configuration is the most stable.
On the other hand, the moment of inertia increases
since the radius in the equatorial plane increases.
Hence the spin-down is a natural consequence of 
the internal field reconfiguration.
It is interesting to note a suggestion
about prolate white dwarfs (Katz 1989).

In order to understand what is physically going on in our model,
we would like to describe typical physical quantities.
Let us consider a star with a polytropic index $n=1$,
a mass $C_M M_0=1.4 M_{\odot}$, and a radius $R=\alpha_0 \xi_0=10^6$ cm.
The central density is given by 
$\rho_0=2.19\times 10^{15} {\rm g\ cm^{-3}}$ from equations 
(\ref{eq:C_M}) and (\ref{eq:alpha0}),
and the mass conservation is always assured by equations 
(\ref{eq:rho1}) and (\ref{eq:rho2}).
We shall consider the first and final state of the magnetic field 
which is characterized by $\lambda=\lambda_2$ and $\lambda=\lambda_1$
respectively.
The lines of force are shown in Fig.~2, and
the fields vanish at the surface of the star.
For simplicity we assume that the decaying magnetic energy
is much smaller than the released energy, i.e., 
$\eta=\Delta {\cal M}=0$
in equations (\ref{eq:delM}) and (\ref{eq:etadef}).
We shall adopt that the increase in the moment of inertia is
$\Delta {\cal I}_{33}/{\cal I}_{0}=10^{-4}$
since this value is required for the positive period increase of SGR 1900+14.
Then, in the first state, 
$\delta_i=1.89\times 10^{-3}$ from equation (\ref{eq:delI}) and 
Tables 1 and 4,
the central density is $\rho_0+\delta \rho_1=
\rho_0+1.58\times 10^{11} {\rm g\ cm^{-3}}$ 
from equations (\ref{eq:rhoexp}) and (\ref{eq:rho1}),
and the star shape is prolate since 
the moment of inertia tensor
of $11$-component is larger than
that of $33$-component, ${\cal I}_{11}/{\cal I}_{33}=1+3.25\times 10^{-4}$,
from equations (\ref{eq:Ixexp}) and (\ref{eq:Izexp}) and Tables 1 and 4.
In the final state, 
$\delta_f=4.46\times 10^{-4}$,
the central density is 
$\rho_0+\delta \rho_1=\rho_0-2.05 \times 10^{11} {\rm g\ cm^{-3}}$,
and the star shape is prolate since 
${\cal I}_{11}/{\cal I}_{33}=1+1.31\times 10^{-4}$.
In this magnetic reconfiguration,
the star becomes more spherical, 
the magnetic energy $C_E {\cal M}=9.60 \times 10^{49} {\rm ergs}$ 
is assumed to be nearly conserved, and
the released energy is $C_E |\Delta {\cal E}|=1.73 \times 10^{46} {\rm ergs}$
from equation (\ref{eq:delE}) and Table 4.

The greatest difficulty in our model
is the assumption of the sudden internal field reconfiguration.
Thompson \& Duncan (1995) considered that
the internal field may diffuse into a configuration
where an interchange instability is no longer
inhibited by topology (Flowers \& Ruderman 1977),
since the diffusion of the strong magnetic field 
will be efficient (Goldreich \& Reisenegger 1992; Thompson \& Duncan 1996).
We may speculate other possibilities, such as
magnetic reconnection (Thompson \& Duncan 1995, 1996),
unpinning of the magnetic vortices in analogy to the
terrestrial hard superconductors (e.g., Sauls 1989), and so on.
In any cases, there are at least two theoretical problems.
First, we have found that the decaying magnetic energy 
during the reconfiguration has to be less than the burst energy
$\Delta {\cal M}\siml 10^{45}$ ergs, i.e.,
the magnetic energy has to be nearly conserved 
$|\Delta {\cal M}/{\cal M}| \siml 10^{-4}$ (see Section \ref{sec:result}).
Second, it is also necessary for the frozen-in condition to break
between equilibria since each equilibrium has different 
$N_P(x)$ and $N_T(x)$ from our analysis (Woltjer 1959).
These constraints may be too strong to give up our model,
or our model may be too simple.
However, the interior of the neutron star is not yet well understood
so that at present we can not say definitely
whether such magnetic reconfiguration can occur or not.
Therefore we had better await the future observational tests (see below)
before reaching the decision.
Anyways, it is an important fact that there exist different neutron star states
separated by the right energy and the right shift in the moment of inertia.

The first observational signature of our model is a pulsation of $\sim$ ms 
period in the burst profile of the giant flare.
In our model, the released energy will be firstly converted into
the oscillational energy of the neutron star (Ramaty et al. 1980).
Since the damping time of the oscillation is about $\sim 0.1$ s
(see below), a pulsation of $\sim$ ms period 
in the initial pulse of the burst profile is expected.
Indeed, there is the pulsating component with $\sim 23$ ms period
in the hard initial pulse of the March 5 event,
which may be the observational signature of the oscillating neutron star
(Barat et al. 1983; Duncan 1998).
In the August 27 event, the data on the initial pulse
with the fine time resolution are not available
(Hurley et al. 1999; Feroci et al. 1999; Mazets et al. 1999).
Since the oscillations may give us the evidence for
the superstrong magnetic field (Duncan 1998),
it is important to observe the giant flares,
especially the initial pulse, with the fine time resolution.

The second observational signature of our model is
the gravitational wave.
According to our model, the released energy
is related to the magnitude of the period increase.
The predicted energy $\simg 10^{46}$ ergs
in Fig.~3 is somewhat larger
than the observed energy $\simg 10^{45}$ ergs of the gamma-ray.
However, it is interesting to note that 
the observation of the Earth's ionosphere indicates
the presence of an intense initial low energy photon
which carries substantially higher (by a factor of $\sim 9$) total energy.
Furthermore, the nonradial ($p$- and $f$-mode) oscillations,
which will be excited in our model,
will be principally damped by the gravitational waves\footnote{
We can make a rather crude estimate as follows.
The luminosity of the gravitational waves can be estimated 
by the quadrupole formula,
$L_{GW}=({G}/{5 c^5}) 
\left\langle {\dddot{\bari}}_{jk} {\dddot{\bari}}_{jk} \right\rangle
\sim ({G}/{c^5}) \left({{\Delta \bari}/{{\Delta T}^3}}\right)^2 
\sim \delta^2 \left({{c^5}/{G}}\right)\left({{G M}/{c^2 R}}\right)^5$,
where $\Delta \bari \sim \delta M R^2$ is the quadrupole of the neutron star,
and $\Delta T \sim \sqrt{R^3/G M}$ is the free fall time.
We estimate the luminosity of the electromagnetic waves
by the dipole formula.
Since there are relations,
$R^2 H \sim$ const,
$\mu \sim H R^3$,
and $\Delta R \sim \delta R$,
where $\mu$ is the magnetic dipole moment,
we can obtain
$L_{EM}=({2}/{3 c^3})\left|\ddot{\bm \mu}\right|^2
 \sim ({1}/{c^3})\left({{\Delta \mu}/{\Delta T^2}}\right)^2 
\sim \delta^2 (c H^2 R^2)\left({{G M}/{c^2 R}}\right)^2$.
The ratio $L_{EM}/L_{GW}\sim \delta (c^2 R/G M)$ 
is small for typical parameters.
Since the above estimate is too crude,
we need a more precise calculation for 
the energy conversion efficiency
from the neutron star oscillation to the gamma-ray.}
(Ramaty 1980; McDermott, Van Horn \& Hansen 1988; 
Cutler, Lindblom \& Splinter 1990).
Decay of these oscillations by gravitational waves
may account for the duration of the initial pulse, $\sim 0.1$ s
(Ramaty 1980; Lindblom \& Detweiler 1983).
The characteristic gravitational wave amplitude
(Thorne 1987) is given
by $h_c=h \sqrt{n}=h \sqrt{f \tau}=(4 G F \tau/\pi c^3 f)^{1/2}$
where $f$ is the gravitational frequency, 
$\tau$ is the damping time of the oscillation,
$n$ is the number of observed cycles
and $F$ is the gravitational wave flux.
Since the gravitational wave flux can be estimated
from the available energy $E$ 
and the distance $d$ of the source
by $F=\dot E/4 \pi d^2=E/8 \pi d^2 \tau$,
we obtain
\beq
h_c\sim 5\times 10^{-21}
\left({{E}\over{10^{47} {\rm ergs}}}\right)^{1/2}
\left({{f}\over{2 {\rm kHz}}}\right)^{-1/2}
\left({{d}\over{5 {\rm kpc}}}\right)^{-1}.
\eeq
Therefore, if the energy that is released as the gravitational waves
is larger than $\sim 10^{47}$ ergs,
which may be possible in our model from Fig.~3,
the gravitational waves from SGRs within $\sim 10$ kpc
could be detected (de Freitas Pacheco 1997; Mosquera Cuesta et al. 1998)
by the planned interferometers
such as LIGO (Abramovici et al. 1992), VIRGO (Bradaschia et al. 1990)
and LCGT (Kuroda 2000).
The event rate will be about $2/20 \sim 10^{-1} {\rm events/yr}$.

Although the energy source of the giant flares
is the gravitational energy in our model,
this energy may be supplied by the decaying magnetic field
before the giant flares (Katz 1982).
The occurrence of two giant flares in different sources within twenty years
most likely suggests that such events occur in most SGRs once 
every $\sim 100$ years (Hurley 1999).
Therefore, if the SGRs are active for $\sim 10^{4}$ years,
the total released energy as the giant flares
is about $\sim 10^{48} (H/10^{16} {\rm G})^4$ ergs.
This energy can be supplied by the total magnetic energy 
$\sim 10^{49} (H/10^{16} {\rm G})^2$ ergs.

We have implicitly assumed that the magnetic axis is parallel
to the spin axis.
If the magnetic axis is inclined to the spin axis,
the spin-up (not spin-down) may also occur at the giant flares
since the moment of inertia along one axis within the equator
decreases as the energy decreases from Fig.~1.
However, there may be some tendency toward an initial alignment of the 
magnetic and spin axis if the magnetic field is generated by a
post-collapse $\alpha$-$\Omega$ dynamo\footnote{
Note that the pulsations of SGRs can occur even for the aligned rotators,
since the four peaked light curve observed during the August 27 event
and the pulse profile of the persistent X-ray component
suggest that SGRs have the multipolar exterior
magnetic structure (Feroci et al. 2000).} 
(Thompson \& Duncan 1993; Thompson et al. 2000).
The electromagnetic (also gravitational) torque
which brakes the rotation of a neutron star also tends to
align the magnetic axis with the rotational one
on the time scale for rotational braking
(Goldreich 1970; Cutler \& Jones 2001; Melatos 2000).
Furthermore, the angular velocity $\Omega$ decreases (i.e., spin-down)
as long as the inclination angle is less than 46 degrees
when the polytropic index is $n=1$
and the state changes from $\lambda_2$ to $\lambda_1$, for example,
since the change of ${\cal I}_{33}$ is comparable
to that of ${\cal I}_{11}$ from Table 4.
In either case, the magnitude of the period jump is related with 
the released energy through equations (\ref{eq:deltaI}) and (\ref{eq:deltaW}).
The misalignment of the spin and the magnetic axis
may also give rise to the modulation of the spin-down history
due to free precession and radiative precession
(Melatos 1999; Thompson et al. 2000), which would provide a direct measure of 
the internal magnetic field.

\section*{Acknowledgments}
We would like to thank T. Nakamura and H. Sato for continuous
encouragement and useful discussions. 
We are also grateful to F. Takahara, T. Shibazaki, S. Shibata,
T. Tsuribe, T. Harada, K. Taniguchi and K. Omukai
for useful discussions,
and T. Nakamura for a careful reading of the manuscript.
This work was supported in part by
Grant-in-Aid for Scientific Research Fellowship (No.9627)
of the Japanese Ministry of Education,
Science, Sports and Culture.

\begin{appendix}
\section{THE REDUCTION OF CERTAIN INTEGRALS}\label{sec:relation}
\subsection{Integrals of $\Theta_0$}
With equation (\ref{eq:LEeq}) and the boundary condition
$\Theta_0(\xi_0)=0$, we can show that (e.g., Chandrasekhar \& Lebovitz 1962)
\beqa
\int_0^{\xi_0} \Theta_0^n \xi^2 d\xi&=&-\xi_0^2 \Theta_0'(\xi_0),
\label{eq:tn}
\\
\int_0^{\xi_0} \Theta_0^{n+1} \xi^2 d\xi&=&
{{n+1}\over{5-n}} \xi_0^3 \left[\Theta_0'(\xi_0)\right]^2.
\label{eq:tn+1}
\eeqa
\subsection{Integrals of $\psi_0$ and $\Psi_0$}
We will show the following relations,
\beqa
\int_0^{\xi_0} n \Theta_0^{n-1} \psi_0 \xi^2 d\xi
&=&-\xi_0^2 \left[\psi_0'(\xi_0)-\Psi_0'(\xi_0)\right],
\label{eq:ptn-1}
\\
I_{\psi}\equiv\int_0^{\xi_0} \Theta_0^n \psi_0 \xi^2 d\xi
&=&{{1}\over{3-n}}\left\{\xi_0^3 \Theta_0'(\xi_0) 
\left[\psi_0'(\xi_0)-\Psi_0'(\xi_0)\right]
-\int_0^{\xi_0} \Theta_0^n \Psi_0' \xi^3 d\xi\right\},
\label{eq:ptn}
\\
\int_0^{\xi_0} \Theta_0^n \Psi_0 \xi^2 d\xi&=&
\xi_0^2 \Theta_0' \left[\psi_0(\xi_0)-\Psi_0(\xi_0)\right]
+(1-n) I_{\psi}.
\label{eq:ptn2}
\eeqa
Firstly, we can verify equation (\ref{eq:ptn-1}) by
making use of equation (\ref{eq:psim}) satisfied by $\psi_0$.
Secondly, we can verify equation (\ref{eq:ptn}) 
by performing partial integrals several times as
\beqa
I_{\psi}&=&-{{1}\over{3}}\int_0^{\xi_0}
\left(\Theta_0^n \psi_0\right)' \xi^3 d\xi
={{1}\over{3}}\left\{
-\int_0^{\xi_0} n \Theta_0^{n-1} \Theta_0' \psi_0 \xi^3 d\xi
-\int_0^{\xi_0} \Theta_0^n \psi_0' \xi^3 d\xi\right\}
\nonumber\\
&=&{{1}\over{3}}\left\{
\int_0^{\xi_0} \Theta_0' 
\left[\xi^2 \left(\psi_0'-\Psi_0'\right)\right]' \xi d\xi
+\int_0^{\xi_0}\left(\xi^2 \Theta_0'\right)' \psi_0' \xi d\xi
\right\}
\nonumber\\
&=&{{1}\over{3}}\Biggl\{
\xi_0^3 \Theta_0'(\xi_0)
\left[\psi_0'(\xi_0)-\Psi_0'(\xi_0)\right]
-\int_0^{\xi_0} \left(\xi \Theta_0'\right)' 
\left(\psi_0'-\Psi_0'\right) \xi^2 d\xi
+\int_0^{\xi_0} \left(\xi^2 \Theta_0'\right)' \psi_0' \xi d\xi
\Biggr\}
\nonumber\\
&=&{{1}\over{3}}\Biggl\{
\xi_0^3 \Theta_0'(\xi_0)
\left[\psi_0'(\xi_0)-\Psi_0'(\xi_0)\right]
+\int_0^{\xi_0} \Theta_0' 
\left(\psi_0'-\Psi_0'\right) \xi^2 d\xi
+\int_0^{\xi_0} \left(2\Theta_0'+\xi \Theta_0''\right) \Psi_0' \xi^2 d\xi
\Biggr\}
\nonumber\\
&=&{{1}\over{3}}\Biggl\{
\xi_0^3 \Theta_0'(\xi_0)
\left[\psi_0'(\xi_0)-\Psi_0'(\xi_0)\right]
-\int_0^{\xi_0} \Theta_0 
\left[\xi^2 \left(\psi_0'-\Psi_0'\right)\right]' d\xi
+\int_0^{\xi_0} \left(\xi^2 \Theta_0'\right)' \Psi_0' \xi d\xi
\Biggr\}
\nonumber\\
&=&{{1}\over{3}}\Biggl\{
\xi_0^3 \Theta_0'(\xi_0)
\left[\psi_0'(\xi_0)-\Psi_0'(\xi_0)\right]
+n I_{\psi}
-\int_0^{\xi_0} \Theta_0^n \Psi_0' \xi^3 d\xi
\Biggr\},
\eeqa
with equations (\ref{eq:LEeq}) and (\ref{eq:psim}),
and the boundary condition $\Theta_0(\xi_0)=0$.
Finally, we can verify equation (\ref{eq:ptn2}) 
by performing the above integration in a different way,
\beqa
n I_{\psi}&=&-\int_0^{\xi_0} 
\Theta_0 \left[\xi^2\left(\psi_0'-\Psi_0'\right)\right]' d\xi
=\int_0^{\xi_0} \Theta_0' \left(\psi_0'-\Psi_0'\right) \xi^2 d\xi
=\xi_0^2 \Theta_0' \left[\psi_0(\xi_0)-\Psi_0(\xi_0)\right]
+I_{\psi}-\int_0^{\xi_0} \Theta_0^n \Psi_0 \xi^2 d\xi.
\eeqa

\subsection{Integrals of $\gamma_0$ and $\Gamma_0$}
We will show the following relations,
\beqa
\int_0^{\xi_0} n \Theta_0^{n-1} \gamma_0 \xi^2 d\xi
&=&-\xi_0^2 \left[\gamma_0'(\xi_0)-\Gamma_0'(\xi_0)\right]
-\int_0^{\xi_0} \sum_{m=0}^{\infty} 
{{n(n-1)}\over{2(2m+1)}} \Theta_0^{n-2} (\psi_m)^2
\xi^2 d\xi,
\label{eq:gtn-1}\\
I_{\gamma}\equiv\int_0^{\xi_0} \Theta_0^n \gamma_0 \xi^2 d\xi
&=&{{1}\over{3-n}}\Biggl\{
\xi_0^3 \Theta_0'(\xi_0)
\left[\gamma_0'(\xi_0)-\Gamma_0'(\xi_0)+{\cal Q}_0\right]
-\int_0^{\xi_0} \Theta_0^n \Gamma_0' \xi^3 d\xi
\nonumber\\
&-&\int_0^{\xi_0} \sum_{m=0}^{\infty} {{n}\over{2(2m+1)}}
\Theta_0^{n-1} \psi_m \left[(4-n)\psi_m+2 \xi \psi_m'\right] \xi^2 d\xi
\Biggr\},
\label{eq:gtn}\\
\int_0^{\xi_0} \Theta_0^n \Gamma_0 \xi^2 d\xi&=&
\xi_0^2 \Theta_0' \left[\gamma_0(\xi_0)-\Gamma_0(\xi_0)\right]
+(1-n) I_{\gamma}
-\int_0^{\xi_0} \sum_{m=0}^{\infty}
{{n(n-1)}\over{2(2m+1)}} \Theta_0^{n-1} (\psi_m)^2
\xi^2 d\xi.
\label{eq:gtn2}
\eeqa
Firstly, we can verify equation (\ref{eq:gtn-1}) by
making use of equation (\ref{eq:gammam}) satisfied by $\gamma_0$
with $Q_{l0}=\psi_l/(2l+1)$.
Secondly, we can verify equation (\ref{eq:gtn})
by performing partial integrals several times as
\beqa
I_{\gamma}&=&{{1}\over{3}}
\int_0^{\xi_0} \left(-n \Theta_0^{n-1} \Theta_0' \gamma_0
-\Theta_0^n \gamma_0'\right) \xi^3 d\xi
\nonumber\\
&=&{{1}\over{3}}\Biggl\{
\int_0^{\xi_0} \Theta_0' \left[\xi^2 \left(\gamma_0'-\Gamma_0'\right)\right]'
\xi d\xi
+\int_0^{\xi_0} \left(\xi^2 \Theta_0'\right)' \gamma_0' \xi d\xi
+\int_0^{\xi_0} \sum_{m=0}^{\infty}
{{n(n-1)}\over{2(2m+1)}} \Theta_0^{n-2} \Theta_0' (\psi_m)^2 \xi^3 d\xi
\Biggr\}
\nonumber\\
&=&{{1}\over{3}}\Biggl\{
\xi_0^3 \Theta_0'(\xi_0) 
\left[\gamma_0'(\xi_0)-\Gamma_0'(\xi_0)\right]
-\int_0^{\xi_0} \left(\xi \Theta_0'\right)' 
\left(\gamma_0'-\Gamma_0'\right) \xi^2 d\xi
+\int_0^{\xi_0} \left(\xi^2 \Theta_0'\right)' \gamma_0' \xi d\xi
\nonumber\\
&+&\int_0^{\xi_0} \sum_{m=0}^{\infty}
{{n}\over{2(2m+1)}} \left(\Theta_0^{n-1}\right)' (\psi_m)^2 \xi^3 d\xi
\Biggr\}
\nonumber\\
&=&{{1}\over{3}}\Biggl\{
\xi_0^3 \Theta_0'(\xi_0) 
\left[\gamma_0'(\xi_0)-\Gamma_0'(\xi_0)+{\cal Q}_0\right]
+\int_0^{\xi_0} \Theta_0'
\left(\gamma_0'-\Gamma_0'\right) \xi^2 d\xi
+\int_0^{\xi_0} \left(2 \Theta_0'+\xi \Theta_0''\right) \Gamma_0' \xi^2 d\xi
\nonumber\\
&-&\int_0^{\xi_0} \sum_{m=0}^{\infty}
{{n}\over{2(2m+1)}} \Theta_0^{n-1} \left[\xi^3 (\psi_m)^2\right]' d\xi
\Biggr\}
\nonumber\\
&=&{{1}\over{3}}\Biggl\{
\xi_0^3 \Theta_0'(\xi_0) 
\left[\gamma_0'(\xi_0)-\Gamma_0'(\xi_0)+{\cal Q}_0\right]
-\int_0^{\xi_0} \Theta_0
\left[\xi^2 \left(\gamma_0'-\Gamma_0'\right)\right]' d\xi
+\int_0^{\xi_0} \left(\xi^2 \Theta_0'\right)' \Gamma_0' \xi d\xi
\nonumber\\
&-&\int_0^{\xi_0} \sum_{m=0}^{\infty}
{{n}\over{2(2m+1)}} \Theta_0^{n-1} \psi_m (3\psi_m +2\xi \psi_m') \xi^2 d\xi
\Biggr\}
\nonumber\\
&=&{{1}\over{3}}\Biggl\{
\xi_0^3 \Theta_0'(\xi_0) 
\left[\gamma_0'(\xi_0)-\Gamma_0'(\xi_0)+{\cal Q}_0\right]
+n I_{\gamma}
-\int_0^{\xi_0} \Theta_0^n \Gamma_0' \xi^3 d\xi
\nonumber\\
&-&\int_0^{\xi_0} \sum_{m=0}^{\infty}
{{n}\over{2(2m+1)}} \Theta_0^{n-1} \psi_m 
\left[(4-n)\psi_m + 2\xi \psi_m'\right] \xi^2 d\xi
\Biggr\},
\eeqa
with equations (\ref{eq:LEeq}), (\ref{eq:gammam}) and (\ref{eq:defQm}),
$Q_{l0}=\psi_l/(2l+1)$
and the boundary condition $\Theta_0(\xi_0)=0$.
Finally, we can verify equation (\ref{eq:gtn2}) 
by performing the above integration in a different way,
\beqa
n I_{\gamma}&=&-\int_0^{\xi_0} 
\Theta_0 \left[\xi^2\left(\gamma_0'-\Gamma_0'\right)\right]' d\xi
-\int_0^{\xi_0} \sum_{m=0}^{\infty} {{n(n-1)}\over{2(2m+1)}}
\Theta_0^{n-1} \left(\psi_m\right)^2 \xi^2 d\xi
\nonumber\\
&=&\int_0^{\xi_0} \Theta_0' \left(\gamma_0'-\Gamma_0'\right) \xi^2 d\xi
-\int_0^{\xi_0} \sum_{m=0}^{\infty} {{n(n-1)}\over{2(2m+1)}}
\Theta_0^{n-1} \left(\psi_m\right)^2 \xi^2 d\xi
\nonumber\\
&=&\xi_0^2 \Theta_0' \left[\gamma_0(\xi_0)-\Gamma_0(\xi_0)\right]
+I_{\gamma}-\int_0^{\xi_0} \Theta_0^n \Gamma_0 \xi^2 d\xi
-\int_0^{\xi_0} \sum_{m=0}^{\infty} {{n(n-1)}\over{2(2m+1)}}
\Theta_0^{n-1} \left(\psi_m\right)^2 \xi^2 d\xi.
\eeqa

\subsection{Integrals of the product of $\psi_m$ and $\Psi_m$}
We will show the following relation,
\beqa
&&\int_0^{\xi_0} \sum_{m=0}^{\infty}
{{n}\over{2(2m+1)}} \Theta_0^{n-1} \psi_m \Psi_m \xi^2 d\xi
\nonumber\\
&=&\int_0^{\xi_0} \sum_{m=0}^{\infty}
{{n}\over{2(2m+1)}} \Theta_0^{n-1} \psi_m 
\left[\psi_m+2 \xi \left(\psi_m'-\Psi_m'\right)\right] \xi^2 d\xi
\nonumber\\
&+&{{1}\over{2}} \xi_0^2 \left[\psi_0'(\xi_0)-\Psi_0'(\xi_0)\right]
\left\{\psi_0(\xi_0)-\Psi_0(\xi_0)
+\xi_0 \left[\psi_0'(\xi_0)-\Psi_0'(\xi_0)\right]\right\},
\label{eq:pP}
\eeqa
which is used to show equation (\ref{eq:W012}).
By using equation (\ref{eq:psim}),
the above equation (\ref{eq:pP}) is shown to be equivalent to
\beqa
&&\sum_{m=0}^{\infty} {{1}\over{2m+1}}
\left[{{1}\over{2}} I_{\psi 1;m}-{{m(m+1)}\over{2}} I_{\psi 2;m}
+I_{\psi 3;m}-m(m+1) I_{\psi 4;m}\right]
\nonumber\\
&=&{{1}\over{2}} \xi_0^2 \left[\psi_0'(\xi_0)-\Psi_0'(\xi_0)\right]
\left\{\psi_0(\xi_0)-\Psi_0(\xi_0)
+\xi_0 \left[\psi_0'(\xi_0)-\Psi_0'(\xi_0)\right]\right\},
\label{eq:pP2}
\eeqa
where
\beqa
I_{\psi 1;m}&=&\int_0^{\xi_0} 
\left[\xi^2 \left(\psi_m'-\Psi_m'\right)\right]'
\left(\psi_m-\Psi_m\right) d\xi,
\\
I_{\psi 2;m}&=&\int_0^{\xi_0}
\left(\psi_m-\Psi_m\right)^2 d\xi,
\\
I_{\psi 3;m}&=&\int_0^{\xi_0}
\left[\xi^2 \left(\psi_m'-\Psi_m'\right)\right]'
\left(\psi_m'-\Psi_m'\right) \xi d\xi,
\\
I_{\psi 4;m}&=&\int_0^{\xi_0}
\left(\psi_m-\Psi_m\right)
\left(\psi_m'-\Psi_m'\right) \xi d\xi.
\eeqa
By performing partial integrals,
we can show that
\beqa
I_{\psi 3;m}&=&\xi_0^3 \left[\psi_m'(\xi_0)-\Psi_m'(\xi_0)\right]^2
-I_{\psi 3;m} +
\int_0^{\xi_0} \left(\psi_m'-\Psi_m'\right)^2 \xi^2 d\xi
\nonumber\\
&=&\xi_0^3 \left[\psi_m'(\xi_0)-\Psi_m'(\xi_0)\right]^2-I_{\psi 3;m}
+\xi_0^2 \left[\psi_m(\xi_0)-\Psi_m(\xi_0)\right]
\left[\psi_m'(\xi_0)-\Psi_m'(\xi_0)\right]-I_{\psi 1;m},
\\
I_{\psi 4;m}&=&\xi_0 \left[\psi_m(\xi_0)-\Psi_m(\xi_0)\right]^2
-I_{\psi 4;m}-I_{\psi 2;m}.
\eeqa
Using the above relations
with the boundary conditions for $\psi_m$ in equation (\ref{eq:bcpsim}),
we find that equation (\ref{eq:pP2}) is satisfied,
and hence equation (\ref{eq:pP}) is satisfied.

\subsection{Integrals of $P_0$ and $T_0$}\label{sec:IM1}
We will show the following relation,
\beq
\int_0^{\xi_0} \Theta_0^n \Psi_0' \xi^3 d\xi
={{1}\over{4\pi}} \int_V \left(-{{1}\over{2}} \varpi^2 P_0 \Delta_5 P_0
+{{1}\over{2}} \varpi^2 T_0^2\right) d^3\xi.
\label{eq:M1relation}
\eeq
First, the left hand side of the above equation
can be transformed as
\beqa
&&4\pi \int_0^{\xi_0} \Theta_0^n \Psi_0' \xi^3 d\xi
=4\pi \int_0^{\infty} \Theta_0^n 
\left[{{1}\over{2}} \int_{-1}^{1} {{\partial}\over{\partial \xi}}
N_P(\varpi^2 P_0) d\mu\right] \xi^3 d\xi
\nonumber\\
&=&\int_V \Theta_0^n 
{{\partial}\over{\partial \xi}} \left[N_P(\varpi^2 P_0)\right]
\xi d^3\xi
=\int_V \Theta_0^n N_P'(\varpi^2 P_0)
{{\partial}\over{\partial \xi}}\left(\varpi^2 P_0\right)
\xi d^3\xi
\nonumber\\
&=&-\int_V \left[\Delta_5 P_0 + T_0 N_T'(\varpi^2 P_0)\right]
{{\partial}\over{\partial \xi}}\left(\varpi^2 P_0\right)
\xi d^3\xi
\nonumber\\
&=&-\int_V \Delta_5 P_0
{{\partial}\over{\partial \xi}}\left(\varpi^2 P_0\right) \xi d^3\xi
-\int_V T_0 {{\partial}\over{\partial \xi}}
\left(\varpi^2 T_0\right) \xi d^3\xi,
\label{eq:M1relation2}
\eeqa
where we use equations (\ref{eq:P0}), (\ref{eq:T0}) and
(\ref{eq:NpP}).
Noting that
\beq
\Delta_5 P_0={{1}\over{\varpi}} {{\partial}\over{\partial \varpi}}
\left[{{1}\over{\varpi}} {{\partial}\over{\partial \varpi}}
\left(\varpi^2 P_0 \right) \right]+{{\partial^2}\over{\partial z^2}} P_0,
\eeq
and $\xi (\partial/\partial \xi)_{\mu}
=\varpi (\partial/\partial \varpi)_{z} + z (\partial/\partial z)_{\varpi}$, 
we divide equation (\ref{eq:M1relation2}) into six parts,
\beq
4\pi \int_0^{\xi_0} \Theta_0^n \Psi_0' \xi^3 d\xi
=I_{M1;1}+I_{M1;2}+I_{M1;3}+I_{M1;4}+I_{M1;5}+I_{M1;6},
\eeq
where
\beqa
I_{M1;1}&=&-\int_V {{\partial}\over{\partial \varpi}}
\left[{{1}\over{\varpi}} {{\partial}\over{\partial \varpi}}
\left(\varpi^2 P_0\right)\right]
{{\partial}\over{\partial \varpi}}
\left(\varpi^2 P_0\right) \varpi d\varpi dz d\varphi,
\\
I_{M1;2}&=&-\int_V {{\partial}\over{\partial \varpi}}
\left[{{1}\over{\varpi}} {{\partial}\over{\partial \varpi}}
\left(\varpi^2 P_0\right)\right]
{{\partial}\over{\partial z}}
\left(\varpi^2 P_0\right) z d\varpi dz d\varphi,
\\
I_{M1;3}&=&-\int_V {{\partial^2}\over{\partial z^2}}
\left(\varpi^2 P_0\right)
{{\partial}\over{\partial \varpi}}
\left(\varpi^2 P_0\right) d\varpi dz d\varphi,
\\
I_{M1;4}&=&-\int_V {{\partial^2}\over{\partial z^2}}
\left(\varpi^2 P_0\right)
{{\partial}\over{\partial z}}
\left(\varpi^2 P_0\right) {{z}\over{\varpi}} d\varpi dz d\varphi,
\\
I_{M1;5}&=&-\int_V \varpi^2 T_0
{{\partial}\over{\partial \varpi}}
\left(\varpi^2 T_0\right) d\varpi dz d\varphi,
\\
I_{M1;6}&=&-\int_V \varpi^2 T_0
{{\partial}\over{\partial z}}
\left(\varpi^2 T_0\right) {{z}\over{\varpi}} d\varpi dz d\varphi.
\eeqa
By performing the partial integrals,
we can show that
\beqa
I_{M1;1}&=&-\int_V \varpi^2 P_0 
{{1}\over{\varpi}}{{\partial}\over{\partial \varpi}}
\left[{{1}\over{\varpi}}{{\partial}\over{\partial \varpi}}
(\varpi^2 P_0)\right] d^3\xi,
\\
I_{M1;2}&=&{{1}\over{2}}
\int_V \varpi^2 P_0 
{{1}\over{\varpi}}{{\partial}\over{\partial \varpi}}
\left[{{1}\over{\varpi}}{{\partial}\over{\partial \varpi}}
(\varpi^2 P_0)\right] d^3\xi,
\\
I_{M1;3}&=&0,
\\
I_{M1;4}&=&-{{1}\over{2}}
\int_V \varpi^2 P_0 {{\partial^2}\over{\partial z^2}} P_0 d^3\xi,
\\
I_{M1;5}&=&0,
\\
I_{M1;6}&=&{{1}\over{2}} \int_V \varpi^2 T_0^2 d^3\xi.
\eeqa
Then, the sum, $(4\pi)^{-1} \sum_{m=0}^{6} I_{M1;m}$, gives
the right hand side of equation (\ref{eq:M1relation}),
and hence we have shown equation (\ref{eq:M1relation}).

\subsection{Integrals of $P_1$ and $T_1$}
We will show the following relation
\beqa
\int_0^{\xi_0} \Theta_0^n \Gamma_0' \xi^3 d\xi
+\int_0^{\xi_0} \sum_{m=0}^{\infty} {{n}\over{2m+1}} \Theta_0^{n-1}
\psi_m \Psi_m' \xi^3 d\xi
={{1}\over{4\pi}}\int_V
\left(-{{1}\over{2}} \varpi^2 P_1 \Delta_5 P_0
-{{1}\over{2}} \varpi^2 P_0 \Delta_5 P_1
+ \varpi^2 T_1 T_0\right) d^3\xi.
\label{eq:M2relation}
\eeqa
First, the first term in the left hand side of the above equation
can be transformed as
\beqa
&&4\pi \int_0^{\xi_0} \Theta_0^n \Gamma_0' \xi^3 d\xi
=4\pi \int_0^{\infty} \Theta_0^n 
\left\{{{1}\over{2}} \int_{-1}^{1} {{\partial}\over{\partial \xi}}
\left[\varpi^2 P_1 N_P'(\varpi^2 P_0)\right] d\mu\right\} \xi^3 d\xi
\nonumber\\
&=&\int_V \Theta_0^n 
{{\partial}\over{\partial \xi}}
\left[\varpi^2 P_1 N_P'(\varpi^2 P_0)\right]
\xi d^3\xi
\nonumber\\
&=&\int_V \Theta_0^n 
\left[N_P'(\varpi^2 P_0)
{{\partial}\over{\partial \xi}}\left(\varpi^2 P_1\right)
+\varpi^2 P_1 N_P''(\varpi^2 P_0)
{{\partial}\over{\partial \xi}}\left(\varpi^2 P_0\right)
\right] \xi d^3\xi
\nonumber\\
&=&-\int_V \Biggl\{
\left[\Delta_5 P_0 + T_0 N_T'(\varpi^2 P_0)\right]
{{\partial}\over{\partial \xi}}\left(\varpi^2 P_1\right)
\nonumber\\
&+&\left[\Delta_5 P_1+n \Theta_0^{n-1} \Theta_1 N_P'(\varpi^2 P_0)
+T_1 N_T'(\varpi^2 P_0)+T_0 \varpi^2 P_1 N_T''(\varpi^2 P_0)\right]
{{\partial}\over{\partial \xi}}\left(\varpi^2 P_0\right)
\Biggr\} \xi d^3\xi
\nonumber\\
&=&I_{M2;1}+I_{M2;2}+I_{M2;3},
\label{eq:M2relation2}
\eeqa
where we use equations (\ref{eq:P0})-(\ref{eq:T1}) and (\ref{eq:Np'P}), and
\beqa
I_{M2;1}&=&-\int_V \left[\Delta_5 P_0
{{\partial}\over{\partial \xi}}\left(\varpi^2 P_1\right)
+\Delta_5 P_1
{{\partial}\over{\partial \xi}}\left(\varpi^2 P_0\right)
\right] \xi d^3\xi,
\\
I_{M2;2}&=&-\int_V \left[
T_0 {{\partial}\over{\partial \xi}}\left(\varpi^2 T_1\right)
+T_1 {{\partial}\over{\partial \xi}}\left(\varpi^2 T_0\right)
\right] \xi d^3\xi,
\\
I_{M2;3}&=&-\int_V n \Theta_0^{n-1} \Theta_1 
{{\partial}\over{\partial \xi}}\left[N_P(\varpi^2 P_0)\right]
\xi d^3\xi.
\label{eq:IM2;3}
\eeqa
In a similar manner to the previous section \ref{sec:IM1},
by performing the partial integrals,
we can show that
\beqa
I_{M2;1}&=&-{{1}\over{2}} \int_V
\left(\varpi^2 P_1 \Delta_5 P_0+\varpi^2 P_0 \Delta_5 P_1\right) d^3\xi,
\label{eq:IM2;1*}
\\
I_{M2;2}&=&-\int_V \varpi^2 T_1 T_0 d^3\xi.
\label{eq:IM2;2*}
\eeqa
With equations (\ref{eq:theta1P}) and (\ref{eq:NpP}),
we perform the angular integrations in equation (\ref{eq:IM2;3})
to find
\beq
I_{M2;3}=-4\pi \int_0^{\xi_0} \sum_{m=0}^{\infty} {{n}\over{2m+1}} 
\Theta_0^{n-1} \psi_m \Psi_m' \xi^3 d\xi.
\label{eq:IM2;3*}
\eeq
By combining equations (\ref{eq:M2relation2}), (\ref{eq:IM2;1*}), 
(\ref{eq:IM2;2*}) and (\ref{eq:IM2;3*}),
the proof of equation (\ref{eq:M2relation})
is completed.
\end{appendix}

%
%

\newpage 
\begin{figure}
\begin{center}  
  \begin{tabular}{cc}
    \epsfysize 8cm \epsfbox{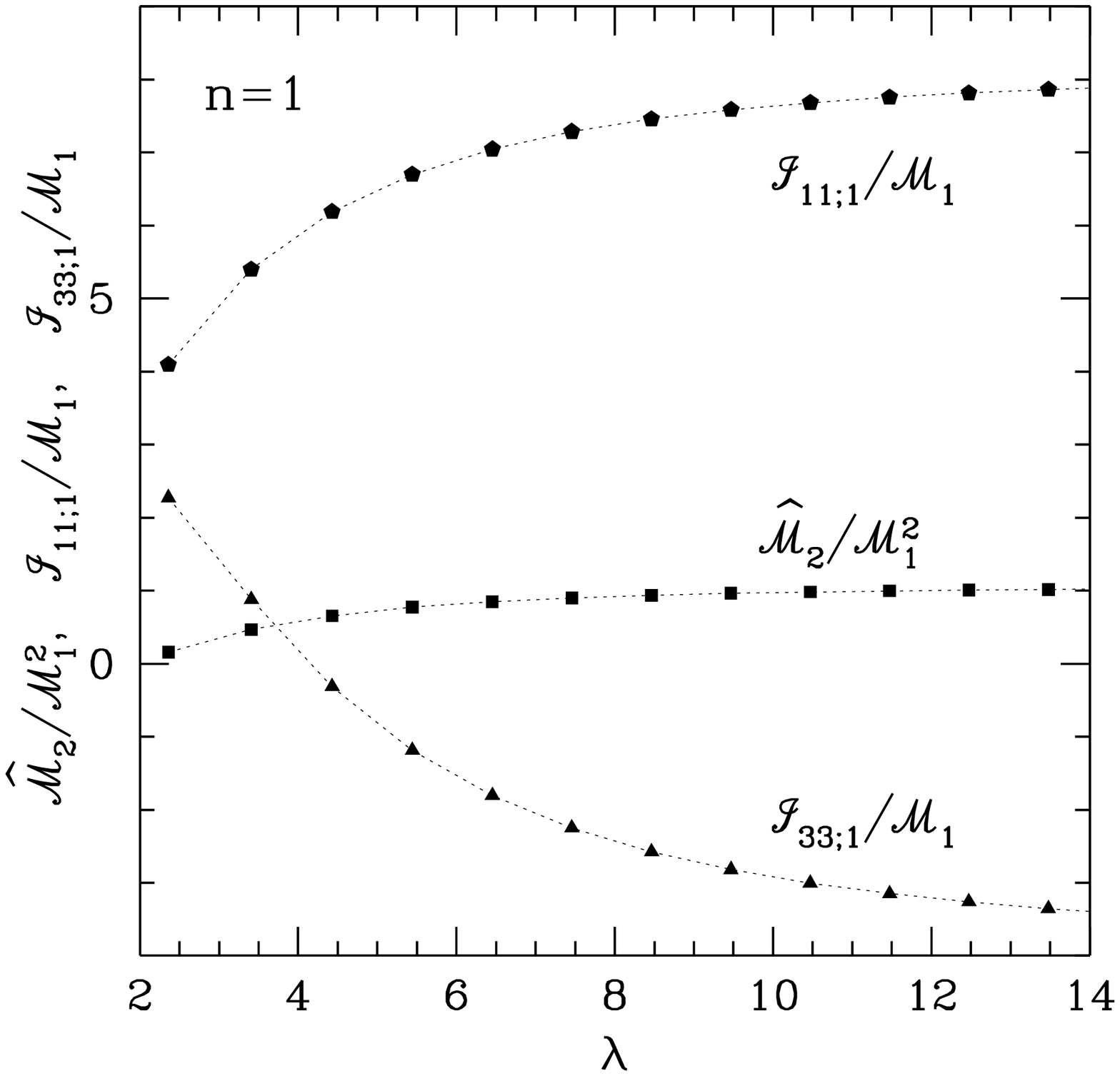} &
    \epsfysize 8cm \epsfbox{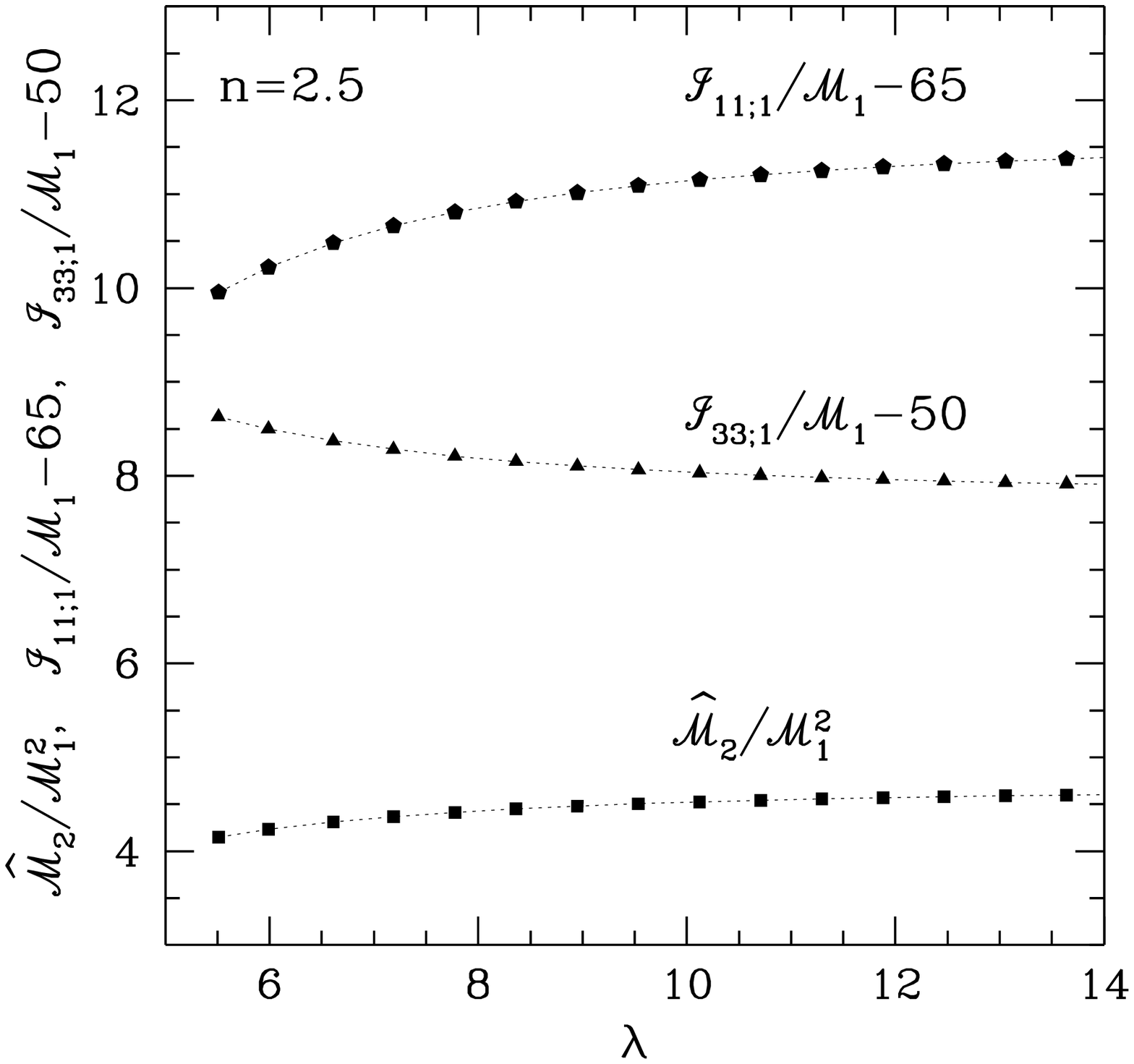}
  \end{tabular}
\end{center}    
\caption[fig1.ps]
{The quantities,
${\widehat {\cal M}_2}/{\cal M}_1^2$, ${\cal I}_{11;1}/{\cal M}_1$ and
${\cal I}_{33;1}/{\cal M}_1$,
in equations (\ref{eq:Econd}) and (\ref{eq:Icond})
are plotted as a function of $\lambda$ for 
the polytropic indices $n=1$ and $n=2.5$.
$\lambda$ controls the configuration of the magnetic fields
and takes the discrete values in Table 1.
As $\lambda$ decreases, 
${\widehat {\cal M}_2}/{\cal M}_1^2$ decreases,
which means that the energy decreases,
${\cal I}_{11;1}/{\cal M}_1$ decreases,
which means that the moment of inertia of $11$-component decreases,
and ${\cal I}_{33;1}/{\cal M}_1$ increases,
which means that the moment of inertia of $33$-component increases.}
\end{figure}

\newpage
\begin{figure}
\begin{center}  
  \begin{tabular}{cc}
    \epsfysize 8cm \epsfbox{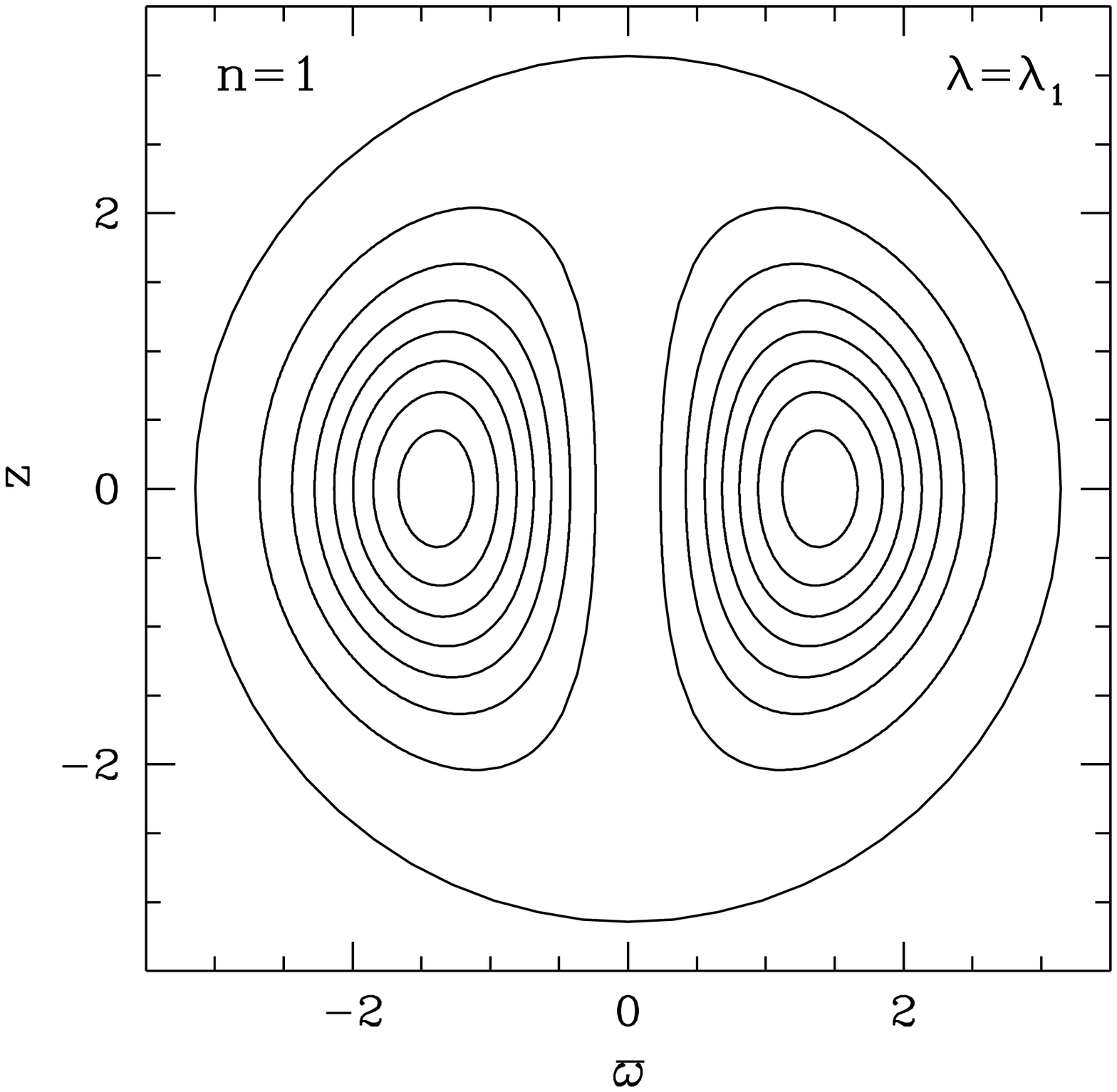} &
    \epsfysize 8cm \epsfbox{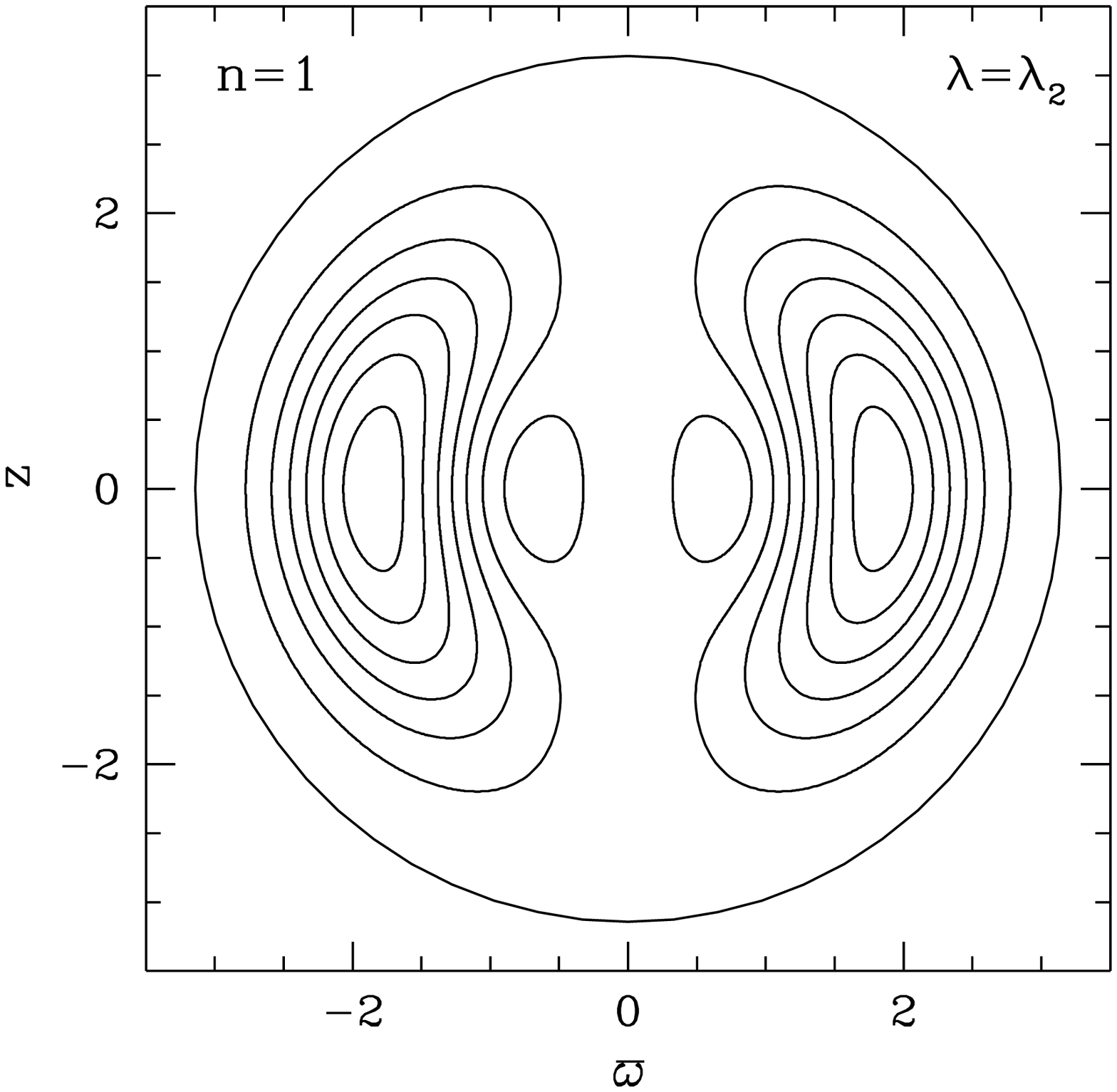}
  \end{tabular}
\end{center}    
\caption[fig2.ps]
{The projection of the lines of force on the meridional
plane, which is given by $\delta^{1/2} \varpi^2 P_0={\rm const}$,
is shown for the case of $n=1$.
$\lambda$ controls the configuration of the magnetic fields
and takes the discrete values in Table 1.
As we can see from this Figure,
the lines of force for larger $\lambda$ have the more
complicated structure than that for smaller $\lambda$.}
\end{figure}

\newpage
\begin{figure}
\begin{center}  
  \begin{tabular}[h]{c}
    \epsfysize 8cm 
    \epsfbox{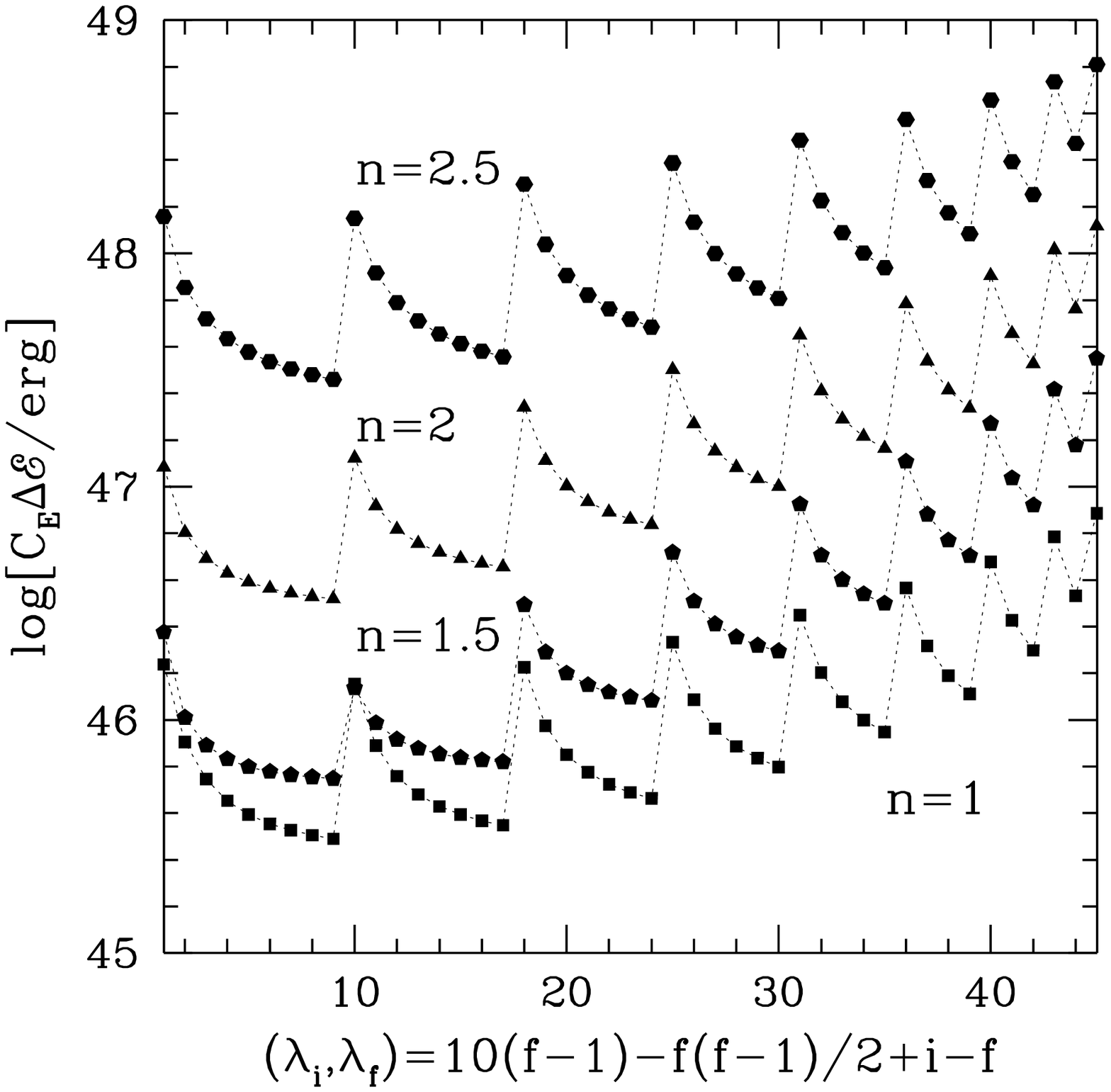}
  \end{tabular}
\end{center}    
\caption[fig3.ps]
{The released energy $C_E \Delta {\cal E}$ 
as a function of 
the various sets of the initial and final state,
$(\lambda_i,\lambda_f)\equiv10(f-1)-f(f-1)/2+i-f$,
where $f=1,\cdots,10,\quad i=f+1,\cdots,10$ and $f<i$
(see Table 2), is shown.
Here we assume $\Delta {\cal I}_{33}/{\cal I}_{0}=10^{-4}$,
which is required for the positive period increase of SGR 1900+14.}

\begin{center}  
  \begin{tabular}[h]{c}
    \epsfysize 8cm 
    \epsfbox{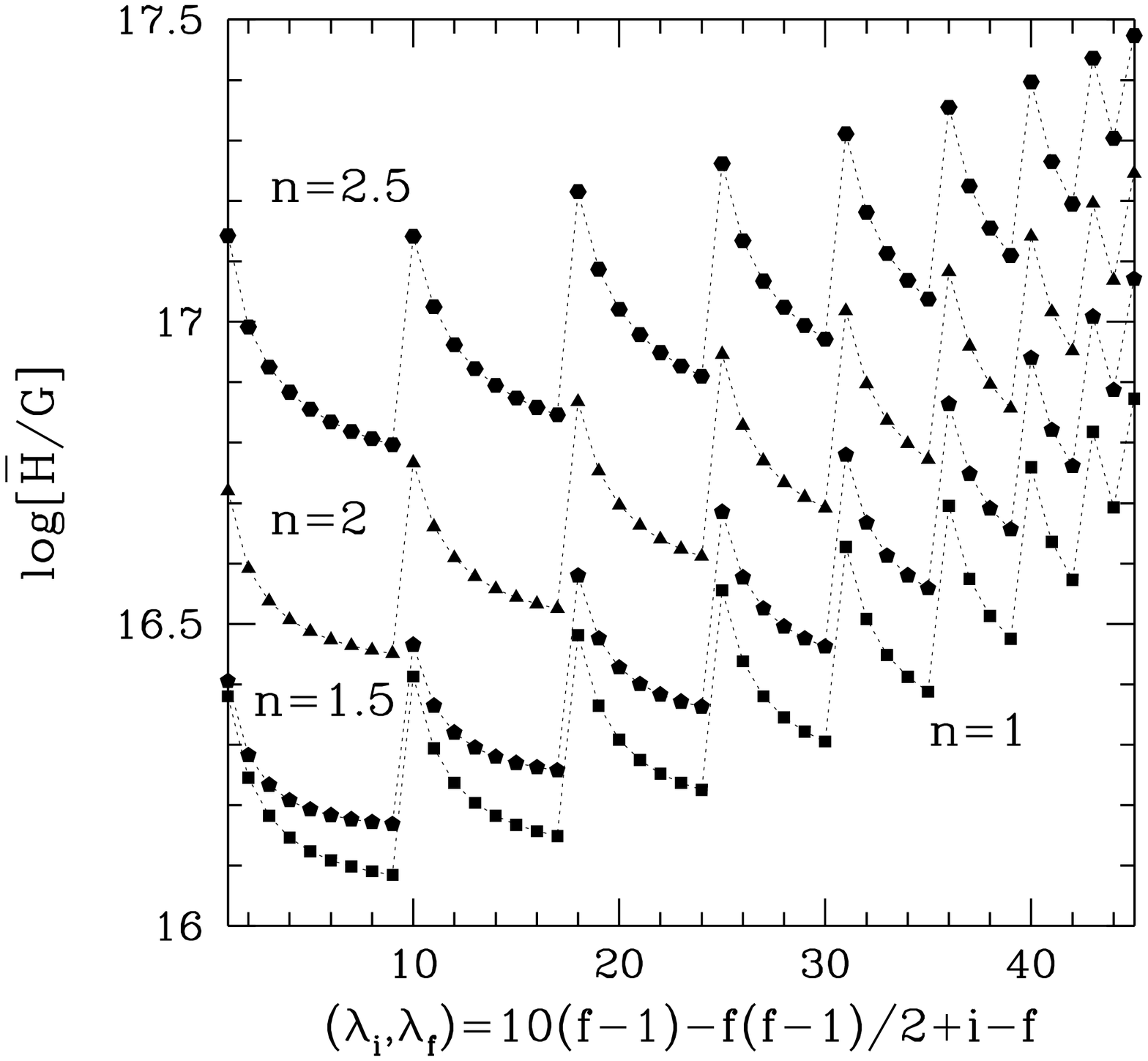}
  \end{tabular}
\end{center}    
\caption[fig4.ps]
{The typical magnetic field $\bar H$,
which is defined by $({\bar H}^2/8 \pi)(4\pi R^3/3)=C_E {\cal M}$,
as a function of 
the various sets of the initial and final state,
$(\lambda_i,\lambda_f)\equiv10(f-1)-f(f-1)/2+i-f$,
where $f=1,\cdots,10,\quad i=f+1,\cdots,10$ and $f<i$
(see Table 2), is shown.
Here we assume $\Delta {\cal I}_{33}/{\cal I}_{0}=10^{-4}$,
which is required for the positive period increase of SGR 1900+14.}
\end{figure}

\newpage
\begin{figure}
\begin{center}  
  \begin{tabular}{cc}
    \epsfysize 8cm \epsfbox{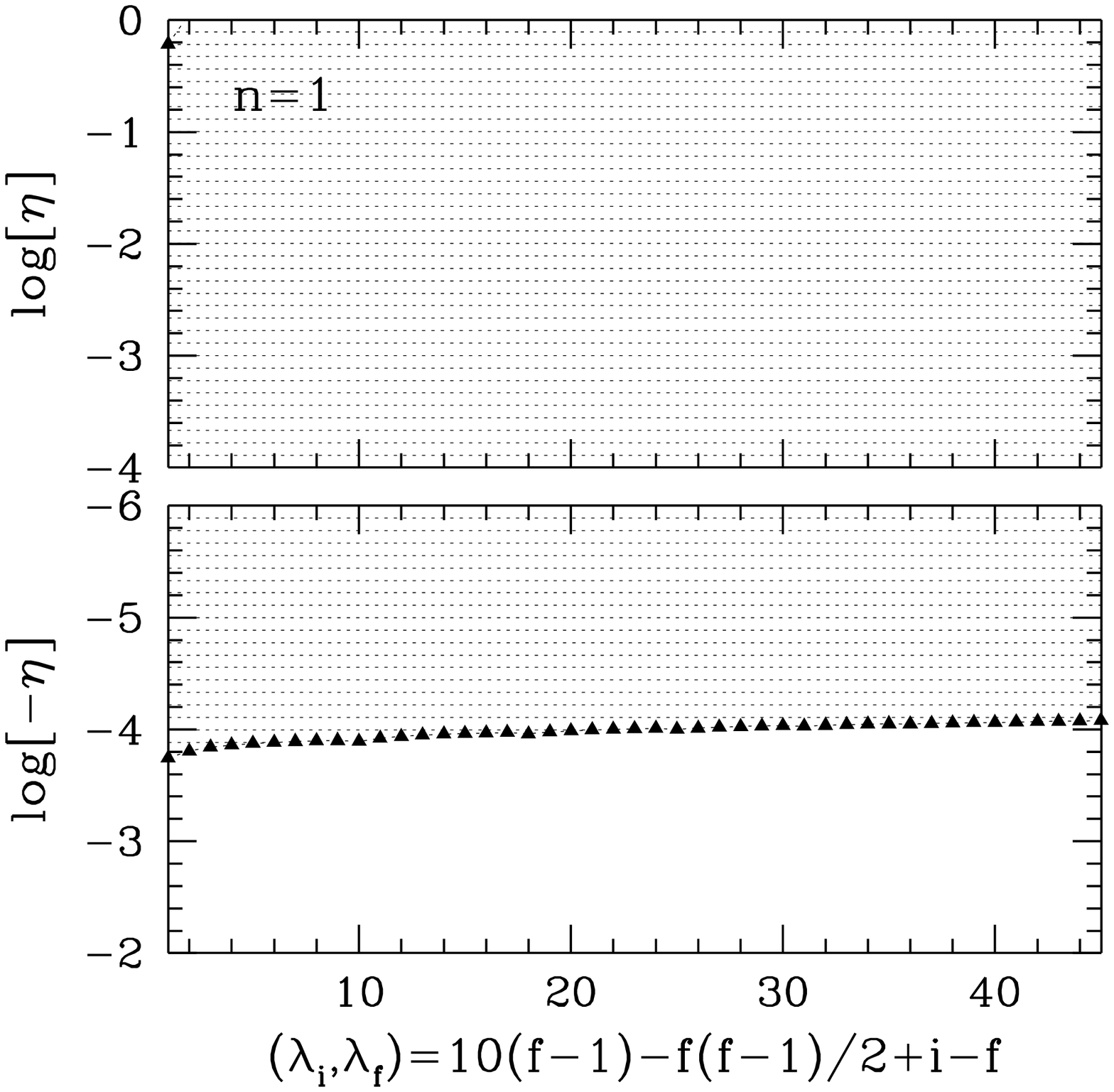} &
    \epsfysize 8cm \epsfbox{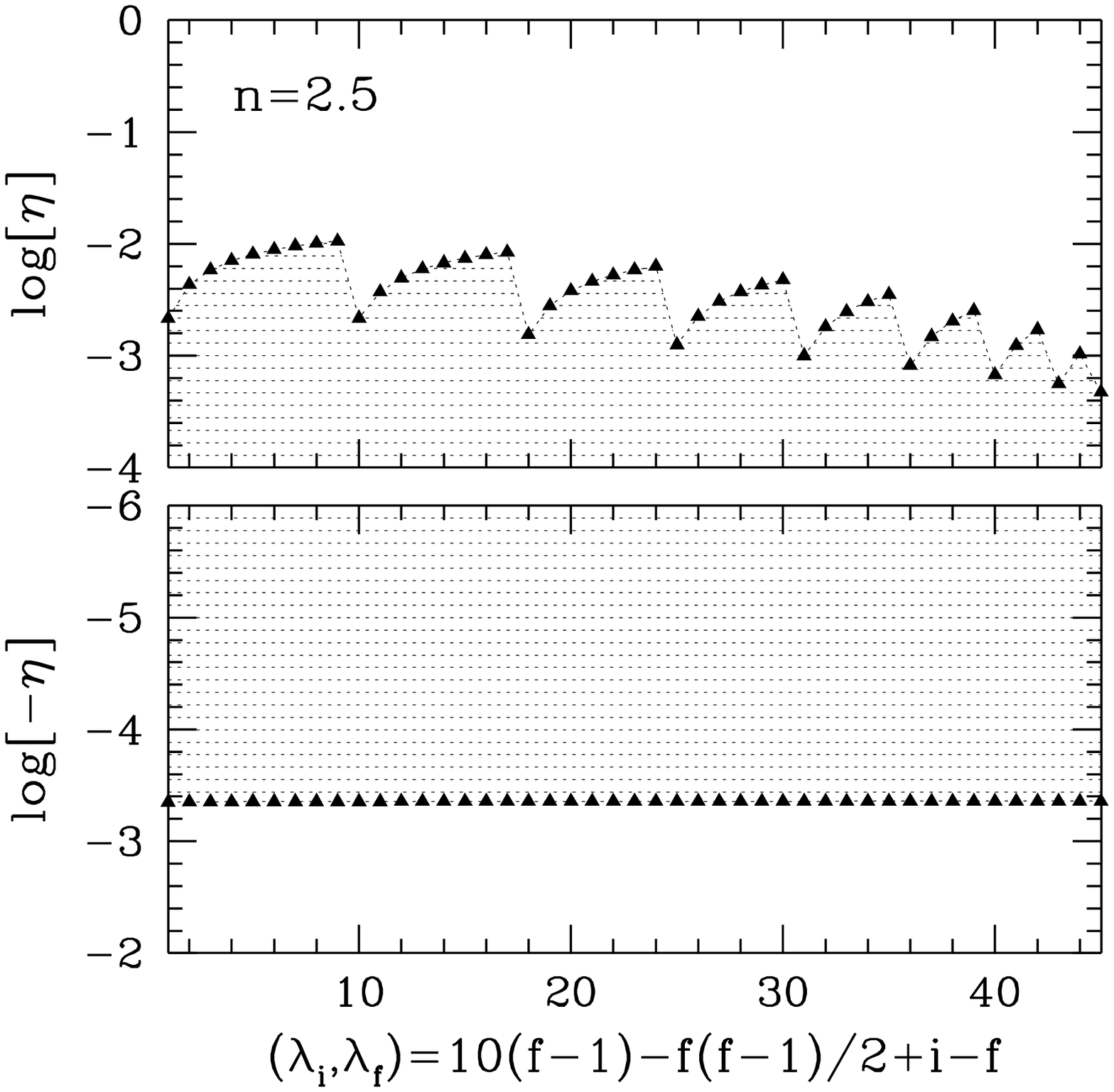}
  \end{tabular}
\end{center}    
\caption[fig5.ps]
{The region of $\eta$ where
the conditions in equations (96) and (97)
are satisfied as a function of
the various sets of the initial and final state,
$(\lambda_i,\lambda_f)\equiv10(f-1)-f(f-1)/2+i-f$,
where $f=1,\cdots,10,\quad i=f+1,\cdots,10$ and $f<i$
(see Table 2), is shown.
$\eta$ denotes the ratio of the decaying magnetic energy 
during the transition to the initial total magnetic energy.
Here we assume $\Delta {\cal I}_{33}/{\cal I}_{0}=10^{-4}$,
which is required for the positive period increase of SGR 1900+14,
and consider the $33$-component of the 
moment of inertia tensor.}
\end{figure}

%
%
\newpage 
\begin{table*}
\begin{center}
\begin {tabular}{ccccc}
  \hline
& $n=1$& $n=1.5$& $n=2$& $n=2.5$\\
\hline
$\lambda_1$ & 2.361933 & 2.449188 & 3.702609 & 5.510036  \\
$\lambda_2$ & 3.407865 & 3.183636 & 4.298552 & 5.988518  \\
$\lambda_3$ & 4.430077 & 4.053194 & 5.047380 & 6.611395  \\
$\lambda_4$ & 5.443462 & 4.915842 & 5.765941 & 7.187845  \\
$\lambda_5$ & 6.452475 & 5.778637 & 6.488668 & 7.776904  \\
$\lambda_6$ & 7.458980 & 6.641027 & 7.210835 & 8.362584  \\
$\lambda_7$ & 8.463904 & 7.503016 & 7.933020 & 8.949183  \\
$\lambda_8$ & 9.467764 & 8.364664 & 8.655152 & 9.535587  \\
$\lambda_9$ & 10.47087 & 9.226033 & 9.377241 & 10.12207  \\
$\lambda_{10}$ & 11.47343 & 10.08718 & 10.09929 & 10.70856 \\
  \hline
\end{tabular}
\end{center}
\caption{The first few roots of $\lambda$
for the polytropic indices $n=1, 1.5, 2$ and $2.5$.}
\end{table*}

\begin{table*}
\begin{center}
\begin {tabular}{cccccccccccccccccccccccc}
  \hline
$(\lambda_i,\lambda_f)$ & 1 & 2 & 3 & 4 & 5 & 6 & 7 & 8 & 9 & 10 & 11 & 12
& 13 & 14 & 15 & 16 & 17 & 18 & 19 & 20 & 21 & 22 & 23 \\
\hline
$i$ & 2 & 3 & 4 & 5 & 6 & 7 & 8 & 9 & 10 & 3 & 4 & 5 & 6 & 7 & 8 & 9 & 10
& 4 & 5 & 6 & 7 & 8 & 9 \\
$f$ & 1 & 1 & 1 & 1 & 1 & 1 & 1 & 1 & 1  & 2 & 2 & 2 & 2 & 2 & 2 & 2 & 2
& 3 & 3 & 3 & 3 & 3 & 3 \\
\hline
$(\lambda_i,\lambda_f)$ & 24 & 25 & 26 & 27 & 28 & 29 & 30 & 31 & 32 & 33 
& 34 & 35 & 36 & 37 & 38 & 39 & 40 & 41 & 42 & 43 & 44 & 45 & \\
\hline
$i$ & 10 & 5 & 6 & 7 & 8 & 9 & 10 & 6 & 7 & 8 & 9 & 10 & 7 & 8 & 9 & 10 
& 8 & 9 & 10 & 9 & 10 & 10 & \\
$f$ & 3 & 4 & 4 & 4 & 4 & 4 & 4 & 5 & 5 & 5  & 5 & 5 & 6 & 6 & 6 & 6 & 7 
& 7 & 7 & 8 & 8 & 9 & \\
\hline
\end{tabular}
\end{center}
\caption{The labelling scheme of the x-axis in Fig. 3, 4 and 5, i.e.,
$(\lambda_i,\lambda_f)\equiv10(f-1)-f(f-1)/2+i-f$
where $f=1,\cdots,10,\quad i=f+1,\cdots,10$ and $f<i$,
is explicitly shown.
$\lambda$ controls the configuration of the magnetic fields
and takes the discrete values in Table 1.
The subscript $i$ ($f$) means the initial (final) state.}
\end{table*}

\begin{table*}
\begin{center}
\begin {tabular}{ccccc}
  \hline
& $n=1$& $n=1.5$& $n=2$& $n=2.5$\\
\hline
$\xi_0$ & $3.14159$ & $3.65375$ & $4.35287$ & $5.35528$  \\
$M_0$ & $3.14159$ & $2.71406$ & $2.41105$ & $2.18720$  \\
$|{\cal E}_0|$ & $1.57080$ &  $8.64015-1$ & $4.45157-1$ & $1.78659-1$  \\
${\cal I}_{0}$ & $8.10448$ & $7.41315$ & $7.07403$ & $7.01315$  \\
\hline
\end{tabular}
\end{center}
\caption{A table of constants for the polytropic indices 
$n=1, 1.5, 2$ and $2.5$. 
$\xi_0$ is the first zero of the Lane-Emden function.
$M_0$ is the normalized mass of the zeroth-order.
$|{\cal E}_0|$ is the normalized total energy of the zeroth-order.
${\cal I}_{0}$ is the normalized moment of inertia tensor
of the zeroth-order.
The number following the plus or the minus sign indicates
the power of 10 by which the table entry should be multiplied.
This notation is followed in all tables.
}
\end{table*}

\newpage
\begin{table*}
\renewcommand{\arraystretch}{1.0}
\begin{center}
\begin {tabular}{ccccccccc}
  \hline
& & ${\cal M}_1$& ${{\widehat {\cal M}_2}/{{\cal M}_1^2}}$&
${{{\cal I}_{11;1}}/{{\cal M}_1}}$&
${{{\cal I}_{33;1}}/{{\cal M}_1}}$&
${{{\cal M}_{T1}}/{{\cal M}_{P1}}}$\\
\hline
$n=1$  &$\lambda_1$
& $1.30707$ & $1.57386-1$ & $4.09418$ & $2.27235$ & $1.86290$ \\
       &$\lambda_2$
& $3.07662-1$ & $4.66603-1$ & $5.39981$ & $8.81646-1$ & $2.94224$ \\
       &$\lambda_3$
& $1.32171-1$ & $6.55343-1$ & $6.18915$ & $-3.14410-1$ & $4.37118$ \\
       &$\lambda_4$
& $7.34761-2$ & $7.73213-1$ & $6.70086$ & $-1.18614$ & $6.15452$ \\
       &$\lambda_5$
& $4.70193-2$ & $8.49530-1$ & $7.04562$ & $-1.80598$ & $8.29340$ \\
       &$\lambda_6$
& $3.28329-2$ & $9.00955-1$ & $7.28585$ & $-2.25076$ & $1.07882+1$ \\
       &$\lambda_7$
& $2.43167-2$ & $9.36929-1$ & $7.45842$ & $-2.57605$ & $1.36392+1$ \\
       &$\lambda_8$
& $1.87852-2$ & $9.62941-1$ & $7.58583$ & $-2.81905$ & $1.68463+1$ \\
       &$\lambda_9$
& $1.49784-2$ & $9.82294-1$ & $7.68219$ & $-3.00433$ & $2.04097+1$ \\
       &$\lambda_{10}$
& $1.22402-2$ & $9.97052-1$ & $7.75663$ & $-3.14830$ & $2.43293+1$ \\
\hline
$n=1.5$& $\lambda_1$ 
& $5.64307-1$ & $4.77567-1$ & $1.02428+1$ & $6.25838$ & $2.42892$ \\
       & $\lambda_2$ 
& $1.77898-1$ & $1.00202$ & $1.23688+1$ & $4.49438$ & $4.66722$ \\
       & $\lambda_3$ 
& $8.48260-2$ & $1.17620$ & $1.32906+1$ & $3.15561$ & $7.78376$ \\
       & $\lambda_4$ 
& $5.08044-2$ & $1.31401$ & $1.37993+1$ & $2.36568$ & $1.18083+1$ \\
       & $\lambda_5$ 
& $3.42333-2$ & $1.40167$ & $1.41011+1$ & $1.87917$ & $1.67192+1$ \\
       & $\lambda_6$ 
& $2.48012-2$ & $1.46078$ & $1.42938+1$ & $1.56431$ & $2.25171+1$ \\
       & $\lambda_7$ 
& $1.88720-2$ & $1.50218$ & $1.44237+1$ & $1.35090$ & $2.92006+1$ \\
       & $\lambda_8$ 
& $1.48800-2$ & $1.53215$ & $1.45153+1$ & $1.20040$ & $3.67684+1$ \\
       & $\lambda_9$ 
& $1.20536-2$ & $1.55447$ & $1.45822+1$ & $1.09066$ & $4.52194+1$ \\
       & $\lambda_{10}$ 
& $9.97399-3$ & $1.57151$ & $1.46325+1$ & $1.00834$ & $5.45528+1$ \\
\hline
$n=2$  & $\lambda_1$ 
& $7.45600-2$ & $1.73936$ & $2.70417+1$ & $1.54539+1$ & $7.51221$ \\
       & $\lambda_2$ 
& $4.92536-2$ & $1.96217$ & $2.77419+1$ & $1.48564+1$ & $1.06080+1$ \\
       & $\lambda_3$ 
& $3.28764-2$ & $2.12083$ & $2.82502+1$ & $1.43747+1$ & $1.49908+1$ \\
       & $\lambda_4$ 
& $2.39851-2$ & $2.22450$ & $2.85612+1$ & $1.40721+1$ & $1.98578+1$ \\
       & $\lambda_5$ 
& $1.83252-2$ & $2.29718$ & $2.87742+1$ & $1.38614+1$ & $2.53951+1$ \\
       & $\lambda_6$ 
& $1.44990-2$ & $2.35003$ & $2.89260+1$ & $1.37100+1$ & $3.15779+1$ \\
       & $\lambda_7$ 
& $1.17785-2$ & $2.38962$ & $2.90381+1$ & $1.35976+1$ & $3.84111+1$ \\
       & $\lambda_8$ 
& $9.76979-3$ & $2.42001$ & $2.91234+1$ & $1.35120+1$ & $4.58946+1$ \\
       & $\lambda_9$ 
& $8.24157-3$ & $2.44384$ & $2.91897+1$ & $1.34452+1$ & $5.40287+1$ \\
       & $\lambda_{10}$ 
& $7.05027-3$ & $2.46286$ & $2.92423+1$ & $1.33922+1$ & $6.28133+1$ \\
\hline 
$n=2.5$& $\lambda_1$ 
& $2.00471-2$ & $4.15142$ & $7.49546+1$ & $5.86265+1$ & $1.79915+1$ \\
       & $\lambda_2$ 
& $1.65192-2$ & $4.23203$ & $7.52209+1$ & $5.85001+1$ & $2.14747+1$ \\
       & $\lambda_3$ 
& $1.32031-2$ & $4.31214$ & $7.54815+1$ & $5.83730+1$ & $2.64277+1$ \\
       & $\lambda_4$ 
& $1.09699-2$ & $4.36892$ & $7.56633+1$ & $5.82826+1$ & $3.14401+1$ \\
       & $\lambda_5$ 
& $9.23785-3$ & $4.41467$ & $7.58082+1$ & $5.82096+1$ & $3.69897+1$ \\
       & $\lambda_6$ 
& $7.89994-3$ & $4.45110$ & $7.59226+1$ & $5.81513+1$ & $4.29372+1$ \\
       & $\lambda_7$ 
& $6.83610-3$ & $4.48080$ & $7.60151+1$ & $5.81038+1$ & $4.93243+1$ \\
       & $\lambda_8$ 
& $5.97686-3$ & $4.50527$ & $7.60909+1$ & $5.80646+1$ & $5.61402+1$ \\
       & $\lambda_9$ 
& $5.27200-3$ & $4.52569$ & $7.61538+1$ & $5.80320+1$ & $6.33882+1$ \\
       & $\lambda_{10}$ 
& $4.68630-3$ & $4.54289$ & $7.62066+1$ & $5.80044+1$ & $7.10677+1$ \\
\hline
\end{tabular}
\end{center}
\caption{A table of integrals for the polytropic indices
$n=1, 1.5, 2$ and $2.5$.
$\lambda_i$, which is given in Table 1,
characterizes the configuration of the magnetic field.
}
\end{table*}

\bsp

\label{lastpage}

\end{document}